\documentclass[iop,apj,twoside,onecolumn,english]{aastex62}

\usepackage{graphicx}
\usepackage{longtable}
\usepackage{amssymb}
\usepackage{multirow}
\usepackage{rotating}

\usepackage{url}
\usepackage{color}
\definecolor{purple}{rgb}{.9,0,.1}
\usepackage{natbib}
\usepackage{lineno}

\begin{document}


\journalinfo{accepted to AJ 20211229}


\newcommand{\kms}{km s$^{-1}$}
\newcommand{\msun}{M$_{\sun}~$}
\newcommand{\rsun}{R$_{\sun}~$}
\newcommand{\lsun}{L$_{\sun}~$}
\newcommand{\lk}{LkH$\alpha$ 225}
\newcommand{\lks}{LkH$\alpha$ 225 South}

\title{LkH$\alpha$ 225 (V1318 Cyg) South in Outburst}

\author{Lynne A. Hillenbrand} 
\affiliation{Department of Astronomy, California Institute of Technology, Pasadena CA 91125, USA}
\email{lah@astro.caltech.edu}

\author[0000-0002-0531-1073]{Howard Isaacson} 
\affiliation{Astronomy Department, University of California, Berkeley, CA 94720, USA}
\affiliation{University of Southern Queensland, Toowoomba, QLD 4350, Australia}

\author[0000-0003-4189-9668]{Antonio C. Rodriguez} 
\affiliation{Department of Physics, Stanford University, Palo Alto, CA 94305-4013, USA}
\affiliation{Department of Astronomy, California Institute of Technology, Pasadena CA 91125}

\author[0000-0002-8293-1428]{Michael Connelley} 
\affiliation{Institute for Astronomy, University of Hawaii at Manoa, 
          640 N. Aohoku Place, Hilo, HI 96720, USA}

\author[0000-0001-8174-1932]{Bo Reipurth}
\affiliation{Institute for Astronomy, University of Hawaii at Manoa, 
          640 N. Aohoku Place, Hilo, HI 96720, USA}

\author[0000-0002-0631-7514]{Michael A. Kuhn}
\affiliation{Department of Astronomy, California Institute of Technology, Pasadena CA 91125, USA}

\author[0000-0002-6881-0574]{Tracy Beck}
\affiliation{Space Telescope Science Institute, 3700 San Martin Drive, Baltimore, MD 21218, USA}

\author{Diego Rodriguez Perez}
\affiliation{Guadarrama Observatory, MPC458, Madrid, SPAIN}

\begin{abstract}
Magakian et al. (2019) called attention to the current bright state of LkHa 225 South, 
a well-known highly embedded, intermediate-mass young stellar object
that over the past two decades has brightened visually from $>20^m$ to $<13^m$. 
We present recent optical photometric monitoring showing colorless, 
non-sinusoidal, periodic brightness oscillations occurring 
every 43 days with amplitude $\sim$0.7 mag.
We also present new flux-calibrated optical and near-infrared spectroscopy, 
which we model in terms of a Keplerian accretion disk, and 
high dispersion spectra that demonstrate similarity 
to some categories of ``mixed temperature" accretion outburst objects. 
At blue wavelengths, LkHa 225 South has a pure absorption spectrum 
and is a good spectral match to the FU Ori stars V1515 Cyg and V1057 Cyg. 
At red optical and infrared wavelengths, however, the spectrum is more similar to Gaia 19ajj, 
showing emission in TiO, CO, and metals. 
\ion{Sr}{2} lines indicate a low surface gravity atmosphere. 
There are also signatures of a strong wind/outflow. 
LkHa 225 South was moderately bright in early the 1950's as well as in the late 1980's, 
with evidence for deep fades during intervening epochs. 
The body of evidence suggests that LkHa225 South is another case of a source 
with episodically enhanced accretion that causes brightening by orders of magnitude, 
and development of a hot absorption spectrum and warm wind. 
It is similar to Gaia 19ajj, but also reminiscent in its long brightening time, 
and brightness oscillation near peak, to the embedded sources L1634 IRS7 and ESO Ha 99.
\end{abstract}
\keywords{Young stellar objects (1834); Circumstellar disks (235); 
Stellar accretion disks (1579); Pre-main-sequence stars (1290); 
}

\section{Introduction}
\begin{figure}
\includegraphics[width=1.00\textwidth]{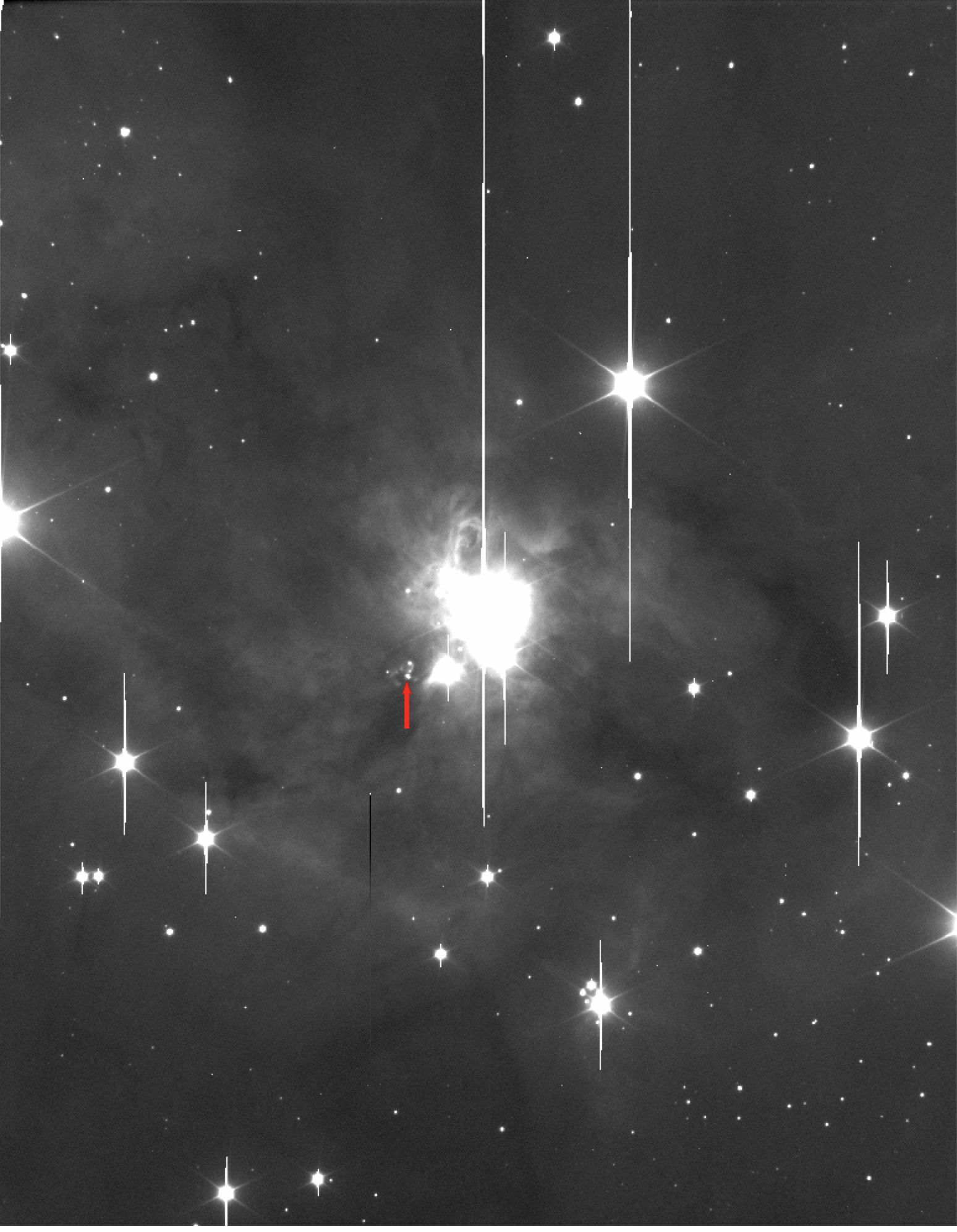}
\caption{
1999 Keck/LRIS $R$-band image ($\sim 6' \times 8'$) 
with orientation north upwards and east to the left.
Bright nebulosity is associated with both
BD+40$^\circ$ 4124 (center) and V1686 Cyg (SE of center).  
The \lk\ North-South pair are due east of V1686 Cyg. 
At this epoch, \lks\ (indicated by red arrow)
was in a prolonged faint state (see Figure~\ref{fig:lc}).  
}
\label{fig:image}
\end{figure}

The young stellar object \lks\ (also designated V1318 Cyg South)
is located at 20:20:30.59 +41:21:26.3 (J2000) and 
is associated with a small cluster of young stars 
usually identified as the BD+40$^\circ$ 4124 cluster.
The two nebulous early type emission-line stars 
BD+40$^\circ$ 4124 (V1685 Cyg) and LkH$\alpha$ 224 (V1686 Cyg)  
that define the optical appearance of the cluster (see Figure~\ref{fig:image})
are long-recognized -- and in fact original -- Herbig Ae/Be stars \citep{herbig1960}. 
But it is \lks\ that is the dominant source in mid-infrared, far-infrared, and
millimeter maps \citep[e.g.][; see also Spitzer and WISE images]{aspin1994}. 
The distance to the region is approximately 900 pc.  

\begin{figure}
\includegraphics[height=2.25truein]{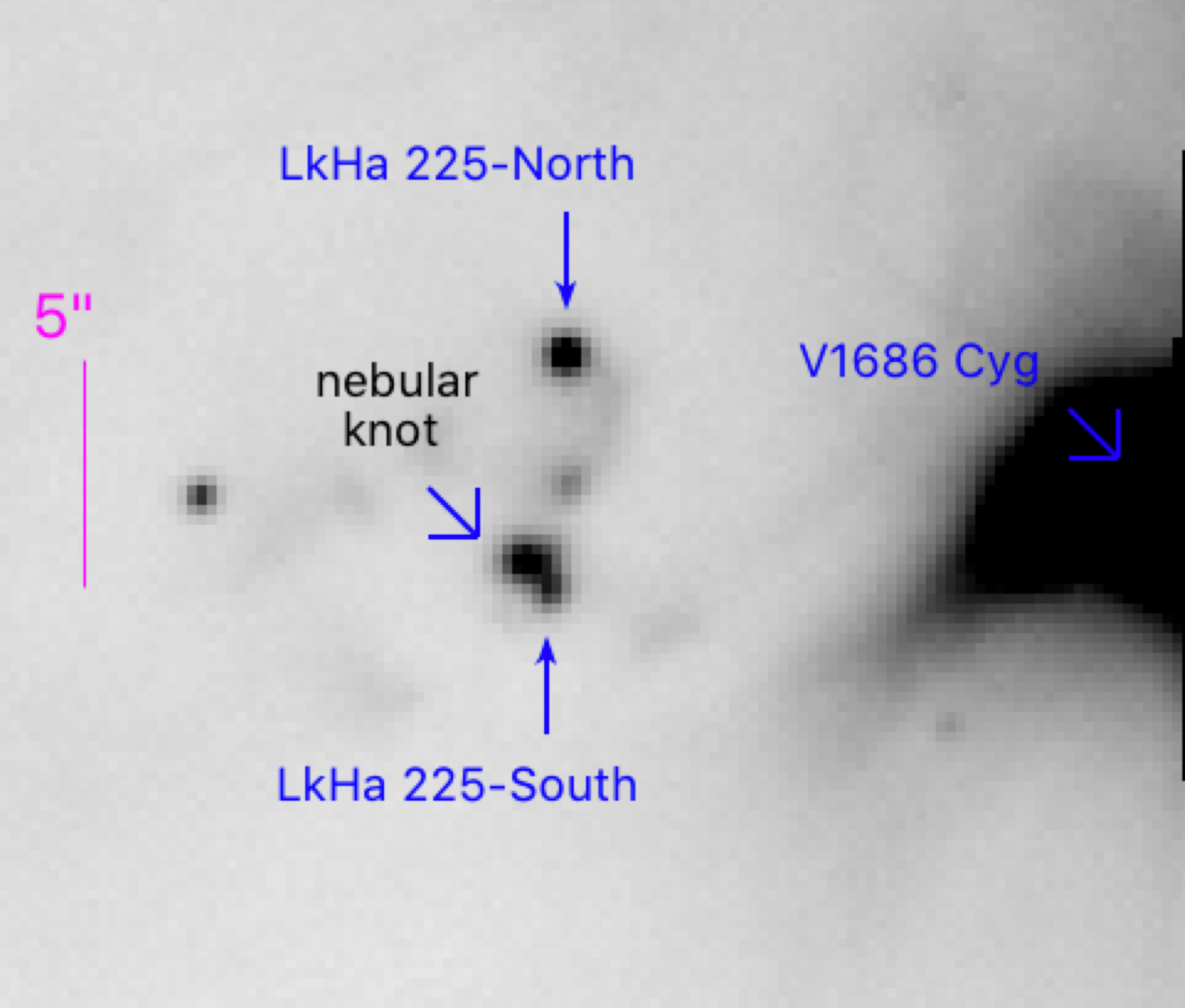}
\\
\includegraphics[height=2.truein]{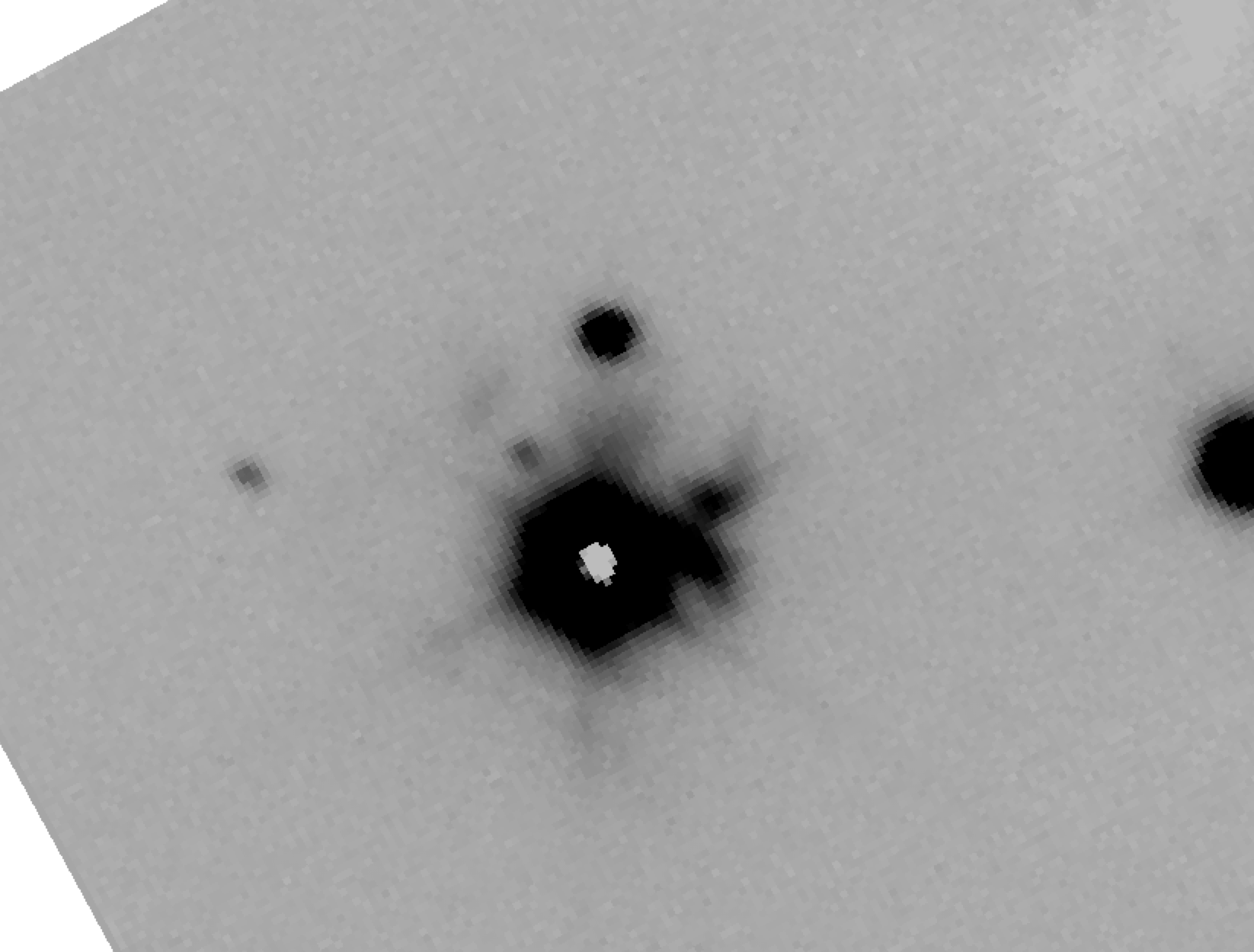}
\caption{
Top panel: Expanded view of the faint-state 1999 $R_c$-band image of 
Figure~\ref{fig:image}, centered here on the \lk\ system, with
\lk\ North and \lks\ both marked. The well-known ridge of nebulosity 
between them is faint in this stretch, but can be seen extending from North 
southward to a nebular knot.  An even brighter nebular knot along the same line,
about 0.8\arcsec\ NE (PA=35.5$^\circ$) of \lks, is marked.  
The position appears to be coincident with maser spots and 
3.1 mm continuum emission; see text.  
The knot is brighter than \lks\ itself at this particular epoch,
and it has some east-west extension.  At the western edge of the image, the
extended bright nebulosity is associated with V1686 Cyg (LkH$\alpha$ 224) 
which is just off frame.
Bottom panel: Bright-state $Y$-band image taken in 2020 with Keck/NIRSPEC-SCAM; 
white pixels indicate saturation in the individual 0.655 sec exposures.
The frame stack comprises 30 seconds of total integration.  
Relative to the earlier image above,
\lks\ has brightened dramatically compared to \lk\ North. 
A faint instrumental ghost appears to the NE of \lks;
this is not the same position as the nebular source 
marked in the top panel.  
}
\label{fig:imagez}
\end{figure}

\begin{figure}
\includegraphics[width=1.00\textwidth]{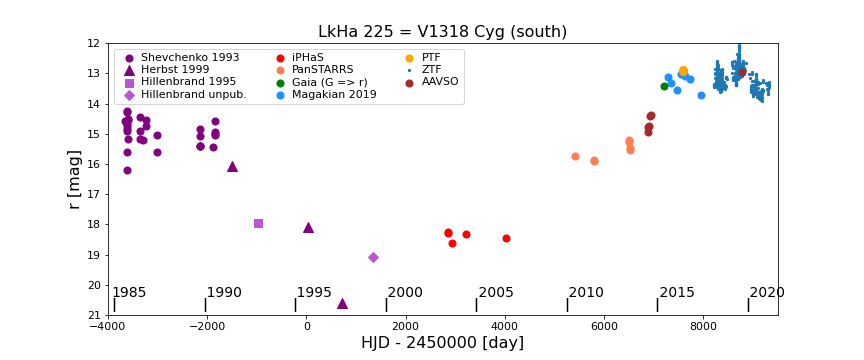}
\caption{
Light curve for \lks\ from the 1980's to the present, demonstrating
the gradual fade over about 15 years and subsequent rise by $>6$ mag 
over the next 20 years.
The published $R$-band data has been transformed to $r$ using an advertised
transformation equation\footnote{\url{http://www.sdss3.org/dr8/algorithms/sdssUBVRITransform.php\#Lupton2005}}, with the correction here typically a shift by 0.6 mag fainter.
The $PanSTARRS$ and ZTF photometry has been converted from 
native AB to Vega magnitudes.
The $Gaia$ DR2 G-band photometry has been corrected into the $r$-band 
using the transformation equation derived by \cite{hillenbrand2018}, 
in this case around 0.1 mag fainter.  
Data points from \cite{herbst1999} are lower limits
and based on their measured $V$-band brightness and an assumed $V-R>3$ mag,
then using the above conversion to $r$.  
Expansion of part of the $ZTF$ time series 
appears in Figure~\ref{fig:lczoomasas}.
}
\label{fig:lc}
\end{figure}

Historically, \lk\ is considered a potential wide binary, 
with a north-south pair separated by about 5\arcsec\ 
and an adjoining ridge of dense nebulosity. 
Both the northern and southern sources have protostellar Class I type 
spectral energy distributions.

\lks\ is generally touted as the source responsible for much of 
the star formation ``activity" in the overall region,  
and is one of its more well-studied objects. 
In addition to being the brightest source locally at mid-infrared wavelengths,
\lks\ has strong millimeter continuum.
Indeed, it is well-known in the massive star, star formation community 
as a deeply embedded, moderate-luminosity protostellar object 
that drives a multi-component, massive bi-polar outflow.

Outflowing gas was first indicated in the spectrum of \cite{andrillat1976} 
via strong absorption in the \ion{He}{1} 10830 \AA\ line, 
consistent with a wind,
along with strong emission in the \ion{Ca}{2} triplet lines.  
\cite{m1997} reported on a spectrum from 1978 with $H\alpha$ described as
``rather intense and possibly has a P Cyg profile" and Na D as ``distinct". 
Shocked H$_2$ gas in the near-infrared was first detected by \citet{aspin1994}
and later mapped by \citet[][their Figure A90]{navarete2015}. 
\cite{vandenancker2000} studied mid-infrared H$_2$
and also present forbidden-line emission maps. 
\citep{sandell2012} detected the outflow in higher spatial resolution [\ion{C}{2}] profiles.
The molecular outflow was investigated in low-J CO by \cite{palla1995,matthews2007} 
and in high-J CO by \cite{sandell2012}. 
The ionized jet was mapped at radio wavelengths by \citet[][their Figure B15]{purser2021}
who found alignment with the H$_2$ lobes of \cite{navarete2015},
and calculated a bolometric luminosity six times that of nearby BD+40$^\circ$ 4124, 
which, adjusting to the distance we adopt, amounts to nearly 2600 $L_\odot$.

\cite{palla1995} had reported H$_2$O maser emission.
\cite{marvel2005} further characterized the maser spot distribution 
and postulated that two outflow sources are involved.
\cite{looney2006} recognized that the maser position is offset 
to the NE from \lks, and demonstrated its coincidence with an extended 3.1 mm continuum source,
suggesting that there may be yet another embedded protostellar source 
at this position.  \cite{bae2011} reported methanol maser emission in the area. 
The potential for spatial confusion seems to render unclear the
origin of the large-scale outflow, and whether it is
\lks, as often advocated, or the embedded mm source that is coincident with the masers.

There is also significant uncertainty about the mass of \lks.  
It is assumed to be intermediate-to-high mass, based on the integrated luminosity in the SED,
with 1600 $L_\odot$ derived by \cite{aspin1994}. 
A spectral type of Ae-Fe was declared by \cite{hillenbrand1995}
for each of the two components of \lk, but this assessment was from a rather 
low signal-to-noise spectrum with little in the way of absorption lines.
Then there is the spectral type of A4Ve reported by SIMBAD, 
that is attributed to \cite{mora2001}; 
this is clearly a typographical error, as the cited paper does not
contain \lk, only the nearby LkH$\alpha$ 224 (V1686 Cyg) 
which is indeed listed as A4Ve by \cite{mora2001}.  
Furthermore, the observations in the paper were taken
in 1998 when \lk\ was faint (as discussed below)
and thus unlikely to have been a successful spectroscopic target.
We make independent progress on a mass estimate in the context of SED fitting
(section \S7), though do not fare much better in constraining $M_*$ 
than the loose estimates already made from observations.

There are a total of three young stellar objects in the BD+40$^\circ$ 4124 cluster 
having the characteristics of intermediate-mass pre-main sequence stars, 
all located within a small molecular core region only $\sim 0.2$ pc in size. 
An accompanying population of lower mass T Tauri type objects is present
as well, but the region is unusual in its high percentage
of intermediate-mass stars
relative to the sub-solar mass population \citep{hillenbrand1995}. 


Our interest in this region was re-piqued when
\cite{m2019} reported a slowly developing outburst in \lks,
finding that the current bright state was reached in 2015.
These authors recount the variability history of the source and
illustrate recent spectra showing a mix of absorption and emission lines,
plus evidence for outflowing gas.

In the current paper, we conduct an extensive investigation 
of the current bright state of the enigmatic source \lks.
Figure \ref{fig:imagez} shows a recent guider camera image from late 2020. 
We report new photometric monitoring from Palomar/P48/ZTF
that indicates an oscillatory nature to the bright-state.
We also present Palomar/P200/DBSP and Keck/HIRES optical spectroscopy, and
IRTF/SpeX and Keck/NIRSPEC infrared spectroscopy, all showing strong
accretion and outflow signatures.  Given its historically faint state,
we have not previously had the opportunity to perform a good quality 
optical photometric and spectroscopic study of \lks. 
We also assemble the long term light curve of the source, and present
previously unpublished imaging and spectroscopy from the faint state, as further context.
Finally, we model \lks\ as an accretion disk dominated system, quantifying
the outburst accretion rate and the stellar parameters.

\section{Historical Light Curve of \lks}

Figure~\ref{fig:lc} illustrates the light curve that can be
assembled for \lks\ over the past 35 years, including photometry
newly reported here that is highlighted in Figure~\ref{fig:lczoomasas}.

The observed brightness in several different red optical filters has been 
converted to an equivalent $r$-band magnitude for plotting purposes.  
At epochs earlier than those shown in the figure, variability of the \lk\ 
system was well-documented during the 1950's to 1970's, as discussed also in detail 
by \cite{m2019}, but the variations were not well-quantified in the literature.
Furthermore, due to the small separation ($\sim$5\arcsec) of the north-south 
pair, the intervening ridge of nebular material, 
and the likely variability of {\it each of} the two components, 
there can be confusion in interpreting the older photometry.
However, the evidence does seem to suggest irregular variations
of at least several magnitudes in \lks. While
typical photographic magnitude brightness was 17-18$^m$, 
two ``eruptions" to approximately 15.5$^m$ were documented by 
\cite{wenzel1972}.  \lks\ can be seen in all available plate data 
from the 1960's as the brighter of the north-south pair 
\citep[e.g. DSS1, ][]{herbig1960,strom1972}. 
However, the source is somewhat fainter in DSS2 images. 

Later, in the 1980's, the \lk\ system was monitored more systematically, 
and magnitudes were reported in known photometric systems.
Typically the brightness measurements were below $16^m$ visually.  
\cite{ibragimov1988} reported large color variations of several magnitudes.
\cite{shevchenko1993} reported $V$-band variations between $15-19^m$ over
JD = 2446345 and 2448159. 
Seemingly inconsistent with the other measurements from this era,
\cite{terranegra1994} report $V=13.5^m$ along with Str\"omgren system photometry
indicating a mid-B spectral type; these data were taken in either\footnote{The observation
dates and the objects are specified, but the correspondence between them is unspecified.} 
September 1988 or June 1990, and are several magnitudes brighter than
the data given in \cite{shevchenko1993}; we disregard them in
what follows, suspecting they may be observations of  
LkH$\alpha$ 224 (V1686 Cyg) just to the west, rather than of LkH$\alpha$ 225.
As discussed above, the same source confusion issues arise 
when considering previously reported spectral types for \lk.

In the early 1990's, \lks\ clearly faded considerably, 
with the $R$-band magnitude increasing from the 14-15$^m$ measured by
\cite{shevchenko1993} to the 17.3$^m$ measured in 1993 
by \cite{hillenbrand1995}.  Based on early 1994 plate data, \cite{m1997}
comment that the source was ``virtually indiscernible even in $I$".
\cite{herbst1999} extended the \cite{shevchenko1993} work to JD=2450710 and
documented continued fading to below $V=24^m$, along with a color increase
in the V=15-19$^m$ mag range from $V-R\approx 1$ to $V-R > 4.5$.  
An additional previously unpublished faint state measurement from 1999
is discussed below. Figure~\ref{fig:imagez} shows that during
this time period, \lks\ was even fainter than \lk\ North.

By the time of the $SDSS$ in 2003, \lks\ had brightened 
to once again become the brighter of the north-south pair, though no $SDSS$ 
photometry is available at the source position \footnote{There is a reported 
measurement at a position within the nebulosity between the 
northern and southern components. In fact, the north-south pair 
plus the connecting nebular ridge is categorized as a galaxy in $SDSS$ catalogs
\citep[as also noticed by][]{m2019}.}.  
The $iPHaS$ measurements reported by \cite{m2019} cover
this missing SDSS epoch, however. 
The subsequent brightening of \lks\ began some time between the last
$iPHaS$ epoch in late 2006 \cite{barentsen2013}
and the first $PanSTARRS$ epoch in mid-2010 \citep{flewelling2020},
as highlighted by \cite{m2019}. 

Finally, a fortuitously-timed set of observations taken by co-author DRP
and reported to AAVSO (as user RZD), captured the photometric rise of \lks\ 
during the period from August 2014 to September 2015.  
This is between the $PanSTARRS$ data acquisition 
and the mean epoch of the $Gaia$ DR2 data point. 
Over the year, \lks\ brightened by about 1 mag in $R$ and reportedly 
also became somewhat redder in color, increasing from $V-R =$ 1.8 to 2.2 mag.
The first \cite{m2019} observation is also in September 2015 and these
authors measure a consistent $V-R=2.08$ mag.  \cite{m2019} document 
color variation at the $\sim0.1-0.2$ mag level over the subsequent two years;
see Figure~\ref{fig:lc}.



\section{Data Assembly Including Old Observations and New Data Acquisition}
\subsection{Imaging and Photometry}

\subsubsection{1999 Keck Imaging Photometry}

$R_c-$ and $I_c-$ band images of the entire BD+40$^\circ$ 4124 cluster including
\lks\ were taken in June 1999
using the Keck Low Resolution Imaging Spectrometer 
\citep[LRIS;][]{oke1995}. These data are illustrated
in Figures~\ref{fig:image} and ~\ref{fig:imagez} and have FWHM of 0.54\arcsec\
sampled at 0.22\arcsec pixel$^{-1}$.
Astrometric calibration and photometry was performed soon thereafter,
as follows.  Within the IRAF environment, point sources 
were identified, centroided, and photometered using a 7 pixel aperture
with a sky annulus extending from 10 to 20 pixels.  This resulted in
measurements for \lk\ North of $R_c=19.40\pm0.01$ mag and $I_c=17.60\pm0.01$ mag,
and for \lks\ of $R_c=19.08\pm0.01$ mag and $I_c=17.21\pm0.01$ mag.
However, as can be seen in Figure~\ref{fig:imagez}, the faint state
of \lks\ reveals an even more complicated morphology with an additional, 
previously unappreciated, extension to the northeast.

The composite photometry from the 7 pixel aperture must therefore be decomposed.
We recently re-examined these images to derive that the separation of
the close pair is 3.78 LRIS pixels or 0.83\arcsec\ at PA=$35.5^\circ$ from SE to NW. 
The \lks-NE component is 0.35 mag brighter than \lks\ itself.
Adopting the combined NE$+$SW brightness as the 19.08 mag reported above,
we find that the individual $R_c$-band brightnesses are 
19.66 mag for \lks-NE and 20.02 mag \lks.  As discussed in Section 4,
it seems likely that this NE component is nebular rather than stellar.

\begin{figure}
\includegraphics[width=0.99\textwidth,trim={-1cm 0 0 0},clip]{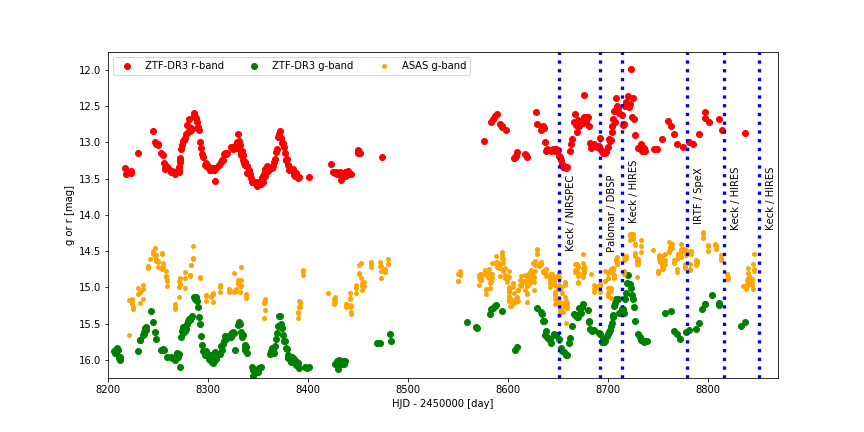}
\caption{
\lks\ time series obtained in the $r$-band and $g$-band 
by $ZTF$ in the 2018 and 2019 seasons; 
error bars are plotted but are generally smaller than the points.  
The source is clearly quasi-periodic in the outburst state, 
with an oscillation timescale of a little more than a month.
The $g$-band photometry reported by $ASAS$ is also included. 
The $\sim$1 mag offset between the ASAS and ZTF $g$-band measurements is likely a 
result of the large ASAS pixels; added variability from \lk\ North 
on top of the quasi-periodic variations exhibited by \lks\ 
perhaps explains the inconsistencies, 
e.g. around MJD$\approx$8350 and 8390 days. 
Epochs of spectroscopy are indicated and labeled.
}
\label{fig:lczoomasas}
\end{figure}

\subsubsection{2018 -- 2020 Optical Survey Photometry}
The Zwicky Transient Facility \citep[ZTF;][]{bellm2019,graham2019} has measured
$g$ and $r$ photometry for \lks\ over three seasons. 
Photometry was harvested from the IPAC/IRSA 
service\footnote{\url{https://irsa.ipac.caltech.edu/cgi-bin/Gator/nph-scan?submit=Select&projshort=ZTF}}
\citep{masci2019} which provides $>$450 measurements in each filter.
$ZTF$ magnitudes are calibrated to $PanSTARRS$  
but do not have color corrections applied.  

Over the same time period as the $ZTF$ dataset, $ASAS$ \citep{shappee2014}  
was also observing the field in the $g$ band. 
As illustrated in Figure~\ref{fig:lczoomasas}, the $ASAS$ light curve mimics 
the $ZTF$ results, though the photometry is $\sim$1 mag brighter. 
The offset between the ASAS and ZTF $g$-band measurements 
is not fully explainable by filter or photometric calibration differences, 
and likely resides in the large photometric aperture of ASAS, 
which would also encompass \lk\ North just 5\arcsec\ away.

\subsubsection{Infrared Photometry Check}

There is unfortunately no infrared monitoring photometry 
available for \lks\ due to the brightness of the source itself
and the crowding of bright sources in this complex region.

In the near-infrared, the J-band data stream coming from 
the Palomar-Gattini-IR survey \citep{de2020} shows that
the source of interest is both saturated and confused.

In the mid-infrared, we similarly come up short on recent monitoring data,
though there is evidence that a brightening has occurred.
\lks\ is saturated in all four bands of the
WISE data products and in the two bands of the NEOWISE reactivation mission.
Furthermore, there is a slight positional offset to the west due to 
the nearby V1686 Cyg. 
However, earlier 2004 epoch measurements from Spitzer 
as reported by \cite{gutermuth2009}, indicated 
[3.6]=5.6 mag and [4.5]=4.4 mag for \lks\ and
[3.6]=5.1 mag and [4.5]=4.4 mag for V1686 Cyg.
The existing WISE data \citep{cutri2012} from 2010 provide only lower limits 
on brightness due to saturation with $[3.4]<3.3$ mag and $[4.6]<1.8$ mag.
Although there is contamination by the near-equal brightness V1686 Cyg,
the WISE data indicate significant mid-infrared brightening of \lks\ between 2004 
and 2010.  Furthermore, NEOWISE monitoring data \citep{cutri2015} 
between 2014 and 2018, which is also highly saturated, 
seem to indicate evidence for brightening from [3.4]$<$3.0 to [3.4]$<$1.5 mag.
As all of the WISE/NEOWISE data 
are beyond the limits of any of the derived saturation corrections\footnote{\url{http://wise2.ipac.caltech.edu/docs/release/neowise/expsup/sec2_1civa.html}}, 
we are unable to comment further on the mid-infrared light curve of \lks.

\begin{figure}
\begin{center}
\includegraphics[width=0.8\textwidth,angle=-90]{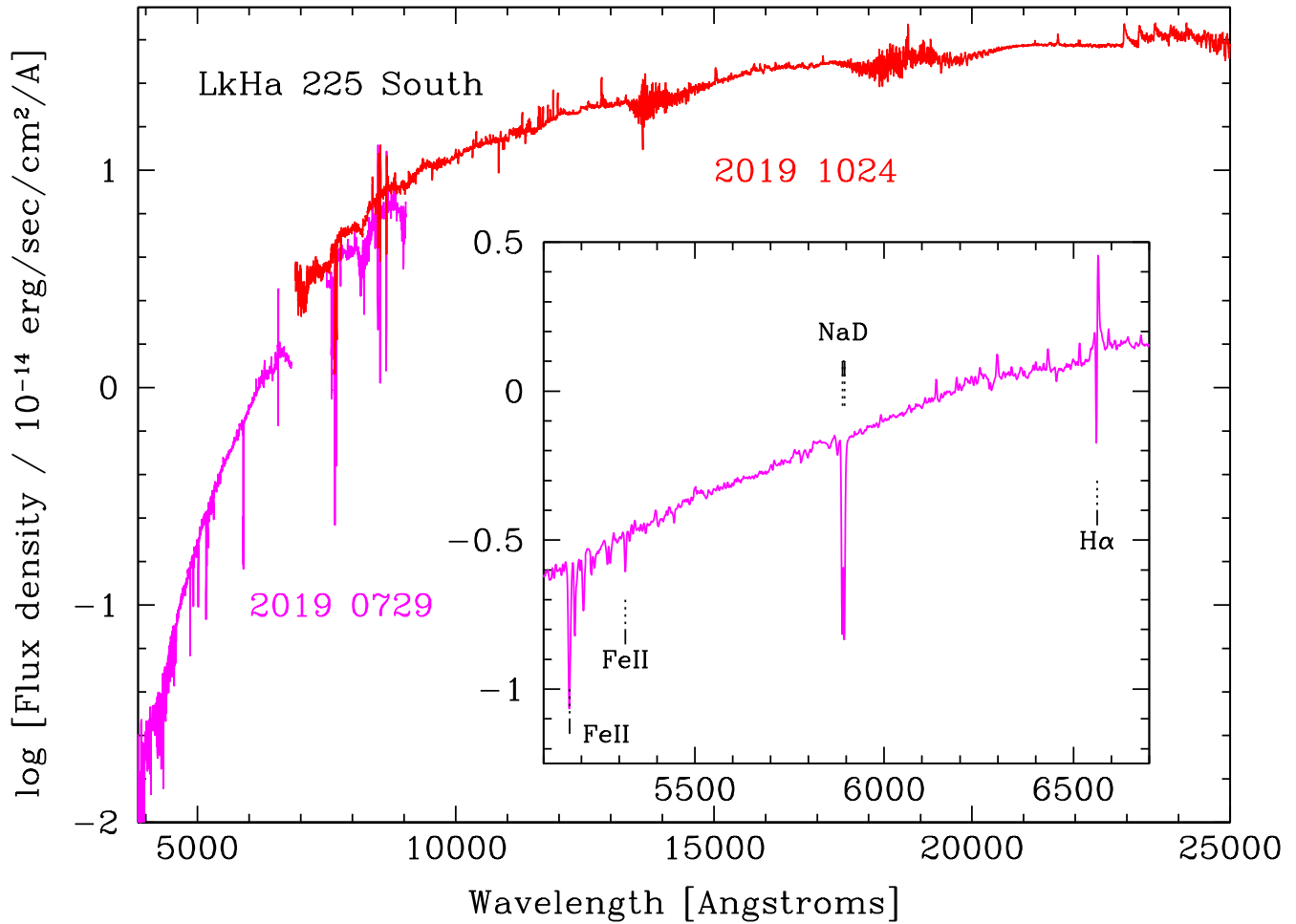}
\end{center}
\caption{
\lks\ low-dispersion optical spectrum from Palomar/DBSP (magenta) and
infrared spectrum from IRTF/SpeX (red).
\lks\ appears significantly reddened, even in outburst. 
Figures \ref{fig:blueandtio} and \ref{fig:spexcompare} respectively 
offer expanded views of the optical and infrared portions of the spectra.
Inset illustrates the strong \ion{Na}{1}D and $H\alpha$ features;
both have blueshifted absorption components when examined in 
higher spectral dispersion data (see Figure~\ref{fig:wind}).
}
\label{fig:spec}
\end{figure}

\begin{figure}
\begin{center}
\includegraphics[width=0.48\textwidth,trim={0.75cm 0 1.75cm 0},clip]{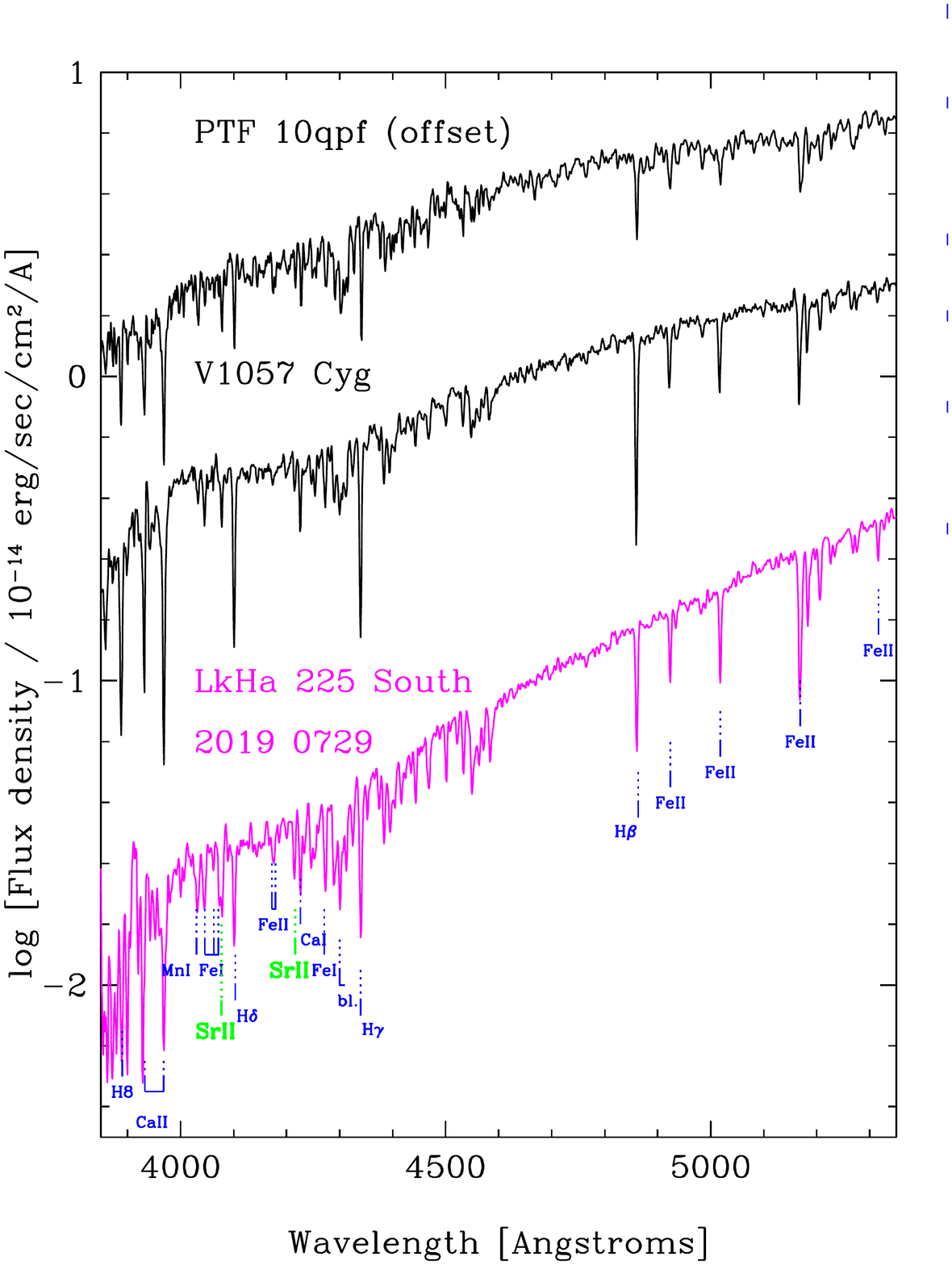}
\includegraphics[width=0.48\textwidth,trim={0.75cm 0 1.75cm 0},clip]{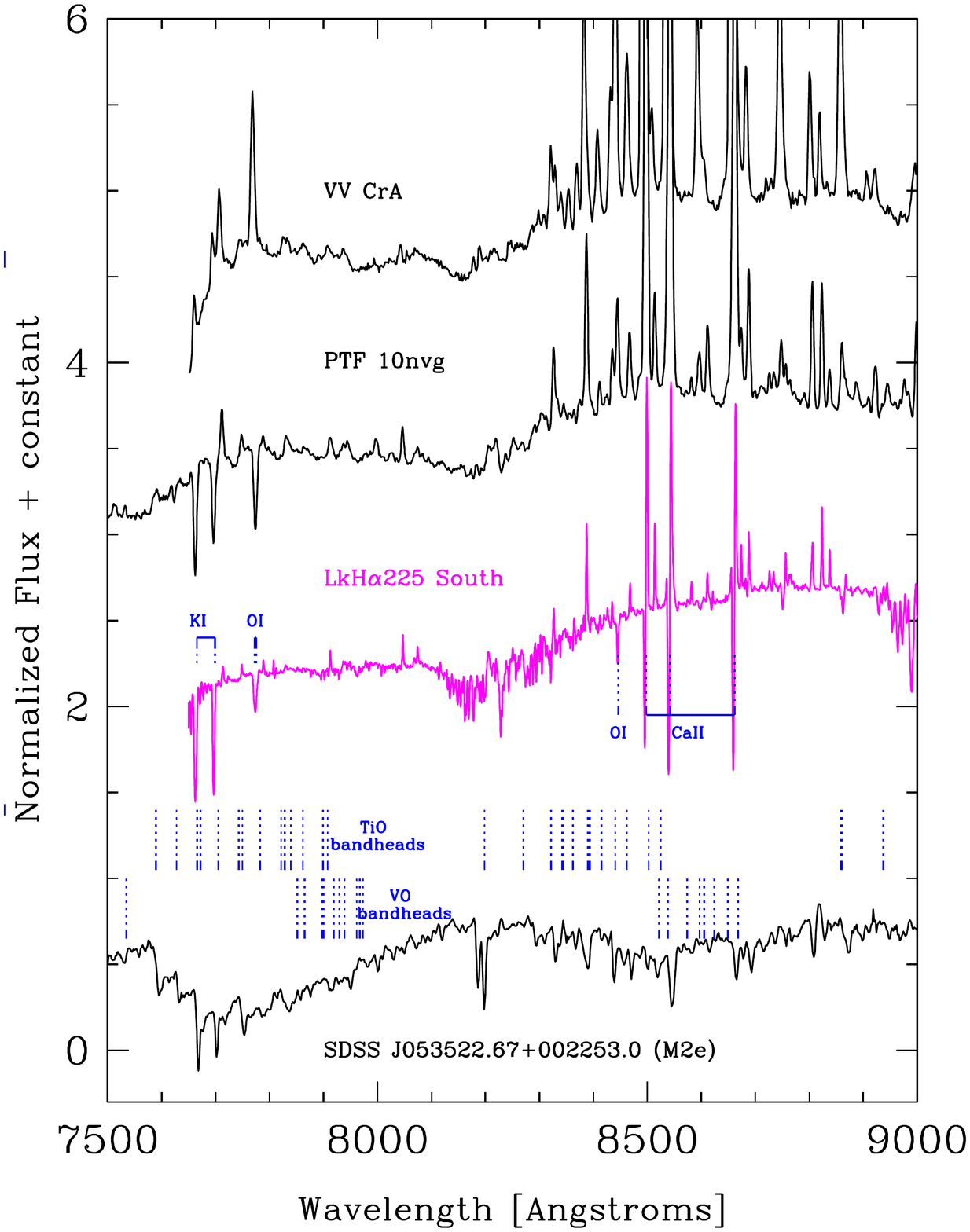}
\end{center}
\caption{
Portions of the Palomar/DBSP optical spectrum of \lks\ (magenta) compared 
to other young stellar objects with rare spectral signatures.
Left panel highlights absorption in \lks\ throughout the blue spectral range, 
with comparison made to the FU Ori stars V1057 Cyg and V2493 Cyg (PTF 10qpf; HBC 722). 
Right panel highlights the prominent TiO emission in the red,
with comparison made to the bright states of 
PTF 10nvg \citep{covey2011} and VV CrA \citep{herczeg2014};
an example field M dwarf with TiO in absorption is also shown.
}
\label{fig:blueandtio}
\end{figure}

\subsection{Spectroscopy}

\subsubsection{Optical}

The venerable Double Spectrograph \citep{og1982} was used at the
Palomar 200" telescope on 27 July, 2019 (UT) to obtain flux-calibrated 
spectra of \lks.  The spectra cover $\sim$3850-6800\AA\ 
with the 600/4000 grating on the blue side, at 1.08\AA/pixel sampling,
and 7500-9000\AA\ with the 1200/7100 grating on the red side,
at 0.40\AA/pixel.  A 1\arcsec\ slit was used with the slit positioned
at the parallactic angle.  Spectra were extracted using the python
package pyraf-dbsp\footnote{\url{https://github.com/ebellm/pyraf-dbsp}}, developed by E. Bellm and B. Sesar, 
as a wrapper to IRAF data processing and spectral extraction tools.  
The realized S/N ratio ranges from $\sim$10 in the far blue 
to $\sim$100 in the far red.  The resulting optical spectrum is illustrated 
in Figures~\ref{fig:spec} and ~\ref{fig:blueandtio}.

We obtained an optical echelle spectrum between $\sim$3400-7900 \AA\ at resolution 
R$\approx$60,000 using the Keck I telescope and HIRES \citep{vogt1994} 
on 18 August, 2019 (UT).
Data acquisition used the standard operating procedures of the 
California Planet Search as described in \cite{howard2010}.  
A 635 sec exposure resulted in a spectrum with S/N = 45 at 5600 \AA\ 
and S/N $\approx$ 150 at 7100 \AA. 

Two additional HIRES spectra were obtained on 
29 November 2019 (UT) and 3 January 2020 (UT),
both covering $\sim$4800-9200 \AA\ at resolution R$\approx$25,000. 
These were processed using the MAKEE reduction pipeline\footnote{\url{https://astro.caltech.edu/~tb/makee/}} 
written by T. Barlow.

\subsubsection{Infrared }

We obtained spectra in the 1 $\mu$m $Y$-band region
at R$\approx$18,500 with the Keck II telescope and the recently upgraded 
\citep{martin2018} NIRSPEC \citep{mclean1998} instrument. 
The 0.576\arcsec\ slit was used with rounds of A-B-B-A position nods 
taken with exposure time of 30 sec per position. 
Spectra were taken on 16 June 2019 (UT) by E. Petigura and T. David,
and on 2 September 2020 (UT) by LAH and J. Spake.
The data were processed using the REDSPEC 
package written by L. Prato, S.S. Kim, \& I.S. McLean. 

We also observed \lks\ over a broader spectral range, covering 0.7-2.4 $\mu$m
at a resolution $R\approx 2000$ 
with the IRTF and SpeX \citep{rayner2003} in its SXD 
(short-wavelength cross-dispersed) mode.
On 24 October 2019 (UT) a set of 15s exposures were taken with 
4 minutes of total exposure time through the 0.3\arcsec\ slit. 
The realized S/N per exposure is $\sim$60 in the 
Y-band and $\sim$200 in the K-band. 

\begin{figure}
\vskip1truein
\includegraphics[width=0.45\textwidth]{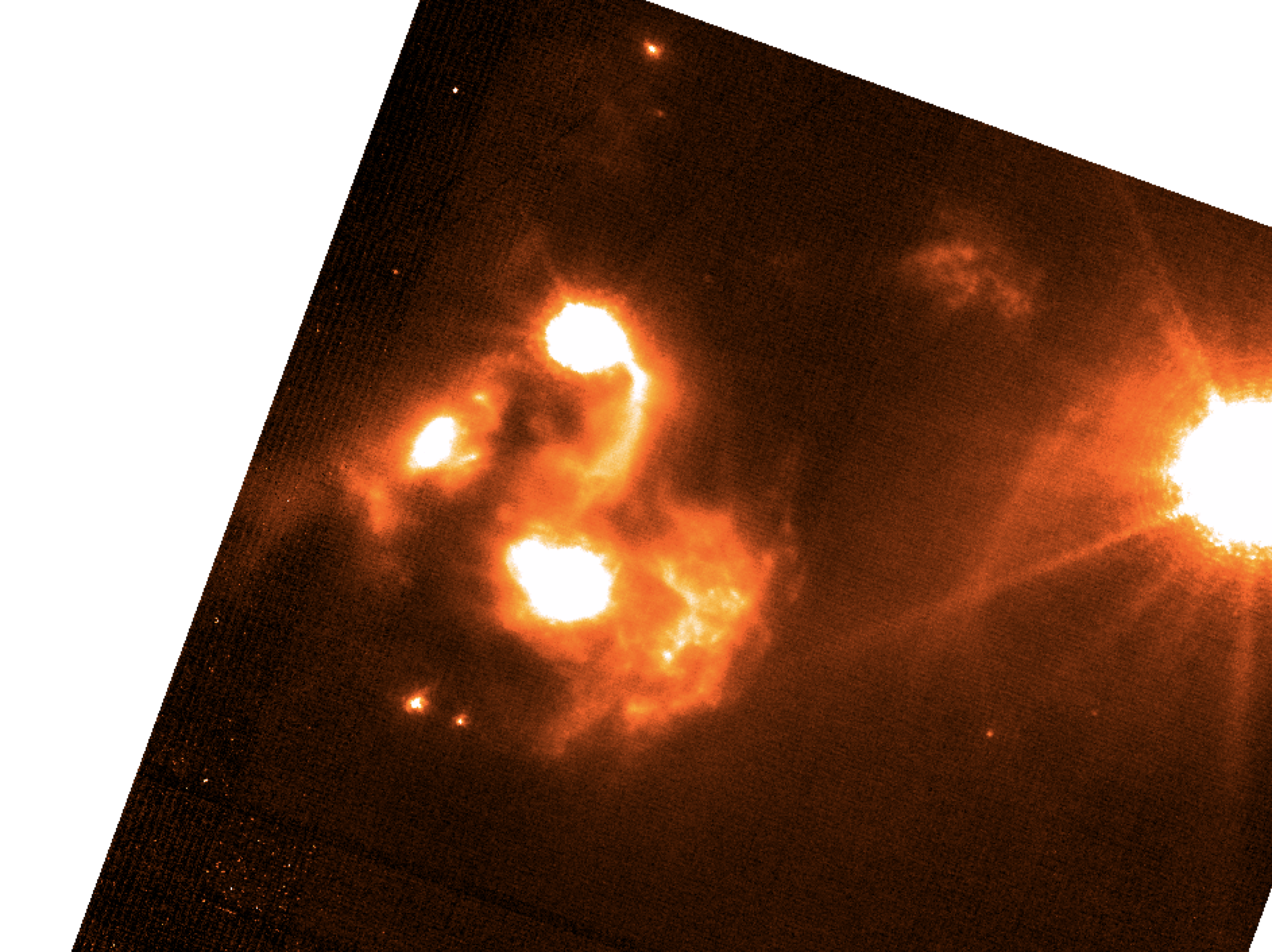}
\hfill
\vskip-4.25truein
\centering
{
\hfill
\includegraphics[width=0.48\textwidth,angle=180]{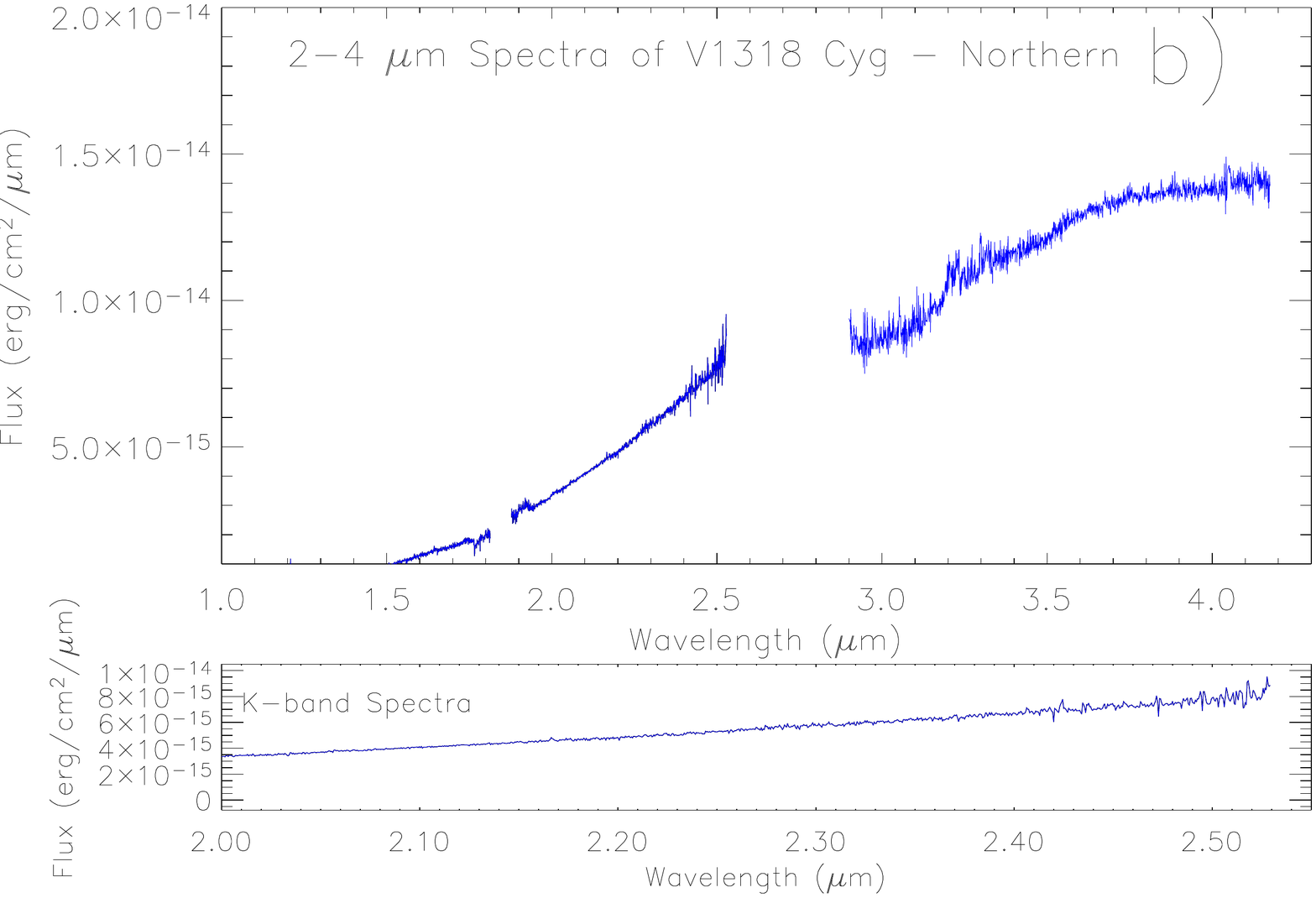}
\\
\hfill
\includegraphics[width=0.48\textwidth,angle=180]{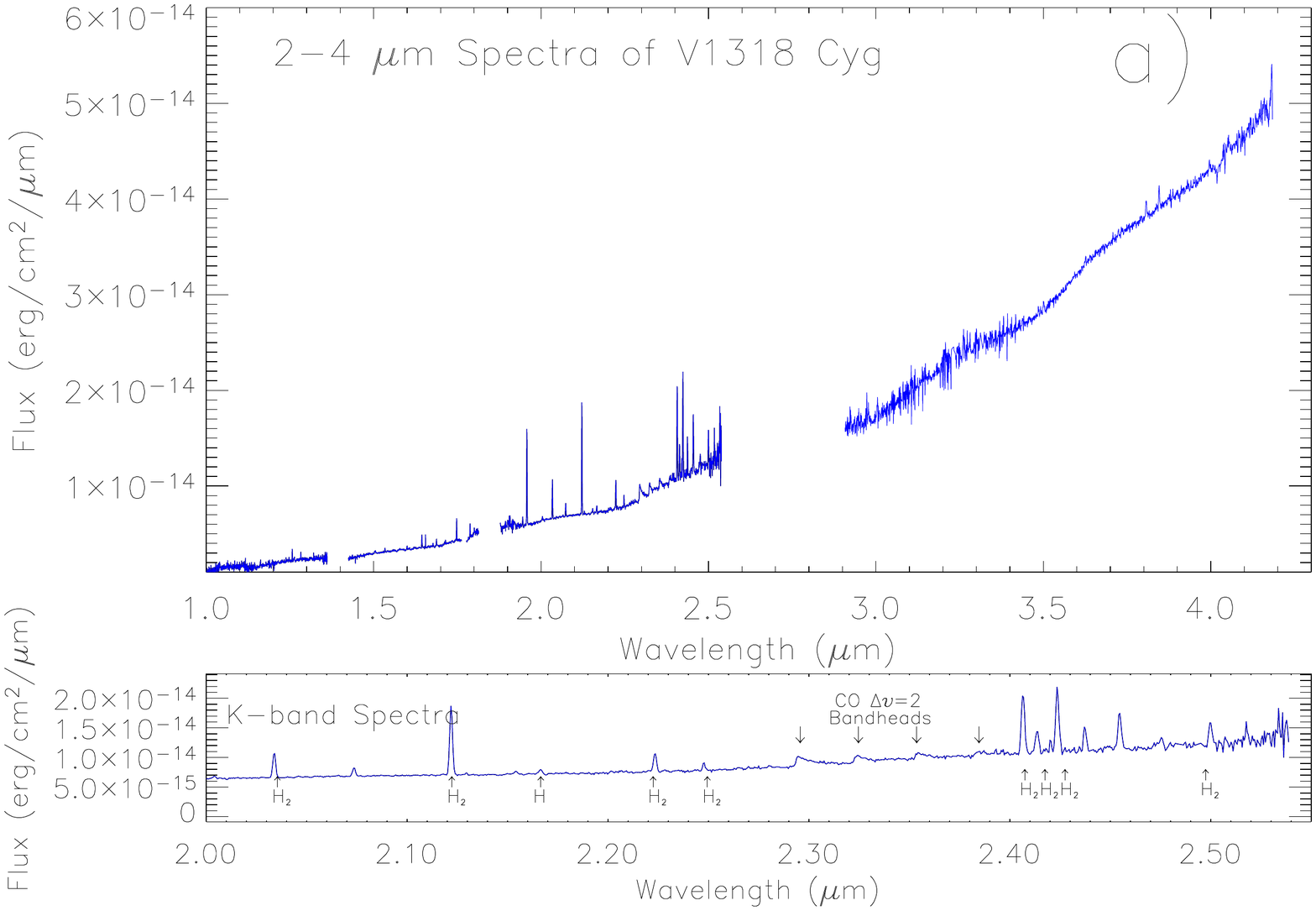}
}
\caption{
Left panel: 
2003 Gemini/NIRI image in H$_2$ 2.12 $\mu$m.
The ridge of nebulosity extending southward from \lk\ North, 
that is apparent in optical images, is present in the near-infrared as well.
Also familiar is the compact nebulosity just northeast of \lks, 
which was seen in our 1999 optical image (Figure~\ref{fig:imagez}). 
Several additional extended nebulosities appear in H$_2$ that are not seen 
in the optical.
Right panels: 
Spectra taken in 2003 of \lk\ North (labeled V1318 Cyg North)
and \lks\ (labeled V1318 Cyg South), during the faint state of the latter. 
The 1-4 $\mu$m data show a red continuum in both components of \lk.
Lower plots in each case highlight the $K$-band spectral region.
}
\label{fig:H2imageGspec}
\end{figure}

Finally, we make use of IRTF/SpeX spectra of \lks\ obtained in a much earlier 
stage of the long term outburst, closer to the faint state in fact. 
On 31 May, 2003 (UT) the 0.5\arcsec\ slit was used to obtain SXD and LXD 
(long-wavelength cross-dispersed)
data
covering 1-4 $\mu$m for both \lks\ and \lk\ North. These spectra are illustrated
in Figure~\ref{fig:H2imageGspec}. Below we compare this older 
\lks\ spectrum to the more recent spectrum taken in the outburst state.  

\section{Small-Scale Morphology of the \lk\ Environment}
The complexity of the region under study was increased by the revelation 
provided in our 1999 Keck/LRIS images that, in its faint state, 
\lk\ consists not only of the well-known wide separation (5\arcsec) 
north-south pair, but an additional spatially resolved component on smaller scales.
This newly appreciated source has a separation of 0.8\arcsec\ 
from \lks\ (Figure~\ref{fig:imagez}). 

Careful consideration of the astrometry shows that the source currently 
outbursting and identified as \lks, specifically as measured by \cite{gaiadr2}, 
coincides with the fainter, SW optical component
of the close pair in Figure~\ref{fig:imagez}.  
The NE component was brighter in 1999, with an east-west extension
that indicates it may be nebular.  

The next available imaging data 
at sufficiently high spatial resolution is a 2003 infrared image from 
the Gemini Science Archive, shown in Figure \ref{fig:H2imageGspec}. 
The SW component is clearly seen as the much brighter source at this epoch,
while the NE component is still apparent.  The H$_2$ emission aspect 
of the NE component can be isolated by subtracting a spatially registered
and scaled continuum image. We do not show such a subtracted image due to 
strong residuals induced by a temporally variable AO PSF,
that affects image alignment and subtraction results. However, we can report
that such a subtraction retains only the extended structure several arcsec 
to the NE and several arcsec to the W-SW of \lks, as well as the very diffuse 
material to the NE of V1686 Cyg. 

The compact nebular component that we are labeling as NE of the point source 
\lks, seems coincident with the location of the maser spot 
field reported by \cite{marvel2005}, as indicated in their Figure 7. 
This is likely also the same position as the 3.1mm continuum peak 
reported by \cite{looney2006}.  These coincidences suggests the possibility of 
an even more deeply embedded companion in the \lk\ system.  Specifically, 
the optical and infrared nebulosity we see near this same position could
be scattered light ($H\alpha$ and H$_2$ emission) 
that escapes along an outflow cone from a protostellar source. 

The physical (projected) source separation between the SW and nebular NE sources
can be calculated using Gaia DR2 \citep{gaiadr2} parallaxes.
The reported parallax of \lks\ itself is rather uncertain, 
and a high renormalized unit weight error is reported with RUWE=8.1, whereas $<$1.4 is recommended. 
The nearby source V1686 Cyg also has a large parallax error. 
We thus use the parallax of the optically brightest source 
in the vicinity, BD+40$^\circ$ 4124 ($1.092\pm0.031$ mas)  
which is better determined.
Adopting 916 pc as the distance to the region, the projected separation 
of NE-SE is 760 AU, while the SW (\lks) to \lk\ North projected separation 
is about 4575 AU.  

Regardless of the nature of this NE component,
we do confirm after considerable astrometric analysis,
that the currently outbursting source is the well-known \lks.

\section{Photometric and Spectro-Photometric Analysis of \lks}

In its current outburst state, \lks\ is red throughout the optical and 
near-infrared wavelength range, as illustrated in Figure~\ref{fig:spec}. 
The overall red continuum slope has a number of contributors,
including: a central stellar source, 
gas and dust emission likely over a range of temperatures, 
and dust extinction from both the circumstellar and the local cloud environment.  
\subsection{Faint State to Bright State Changes}

While \lks\ remains red in an absolute sense,
the optical colors in the outburst state seem somewhat, 
but not dramatically, bluer than those measured much earlier.
The $g-r$ color of 2.8 mag (AB) reported by ZTF 
would correspond to $V-R\approx1.9$ mag and $R-I\approx1.5$ mag.
This is based on a conversion from $g-r$ (AB) to $V-R$ (AB) 
from \cite{jordi2006}, 
further correction from AB to Vega magnitudes,  
and then scaling $V-R$ to $R-I$ according to the relation between these colors 
in the \lks\ monitoring data of \cite{shevchenko1993}. 

The source colors were similar in the periods in the 1980's, 
before the long duration fade to $>20^m$, 
as well as during the fade e.g. 1994 data 
\cite[][; $V-R_c$=1.8: mag and $R_c-I_c$=1.7: mag]{hillenbrand1995},
and within the deep fade, 
e.g. the 1999 Keck/LRIS imaging (\S 3.1; $R_c-I_c=1.87$ mag).

Color information during the re-brightening is scarce. 
The 2014 measurements by co-author DRP that were reported to AAVSO (as user RZD) 
were $V-R=1.7-2.2$ mag. They show the V-R color becoming redder 
as the source brightens by about 0.6 mag.
This is consistent with the fact that the currently measured $g-r = 2.8$ mag 
(AB) is redder than the $PanSTARRS$ values of $g-r = 2.1$ mag (AB), 
measured in 2013, and 2.3 mag (AB), measured in 2011.
The current values are on the blue side of the colors measured 
in 2015 by \cite{m2019}, which also represent the bright state, 
reported as $V-R=2.1$ mag and $R-I$=1.5-1.8 mag. 

We note that for the majority of these photometric measurements, 
the errors are often unreported.  
As a guide to approximate values, we consider that the $g-r$ color
error in the PanSTARRS (PS1) catalog is 0.1 mag, and we take this as a 
minimum error for the other color measurements given above.

Even though the current optical colors do not seem significantly different 
from those measured at much earlier epochs,
the evidence seems to suggest a reddening trend as the source
rose to its current peak brightness.  However, this is not what is seen
in the infrared.  

Although no colors are available, we do have
spectrophotometric measurements.
To determine whether the infrared SED has changed during the brightening, 
we can directly examine the 2003 infrared spectrum 
relative to the 2019 infrared spectrum; both are illustrated in 
Figure~\ref{fig:spexcompare}.  A ratio of the two spectra shows a 
flux increase during the burst by a factor of $\sim100$ in the J-band, 
$\sim$85 in the H-band, and $\sim$50 in the K-band, implying a ``blue" nature to the burst.  

\subsection{Extinction Effects}

There is no value of extinction that would truly flatten the infrared flux ratio
spectrum described above. This implies that the dramatic brightening 
can not be explained by a reduction in extinction alone.
However, a decrease in $A_V$ by about 3 mag would nearly equate the ratio at
$J$-band and $K$-band around a value of $\sim 35$, with the $H$-band a little higher. 
A reasonable scenario is one in which there has been both 
a reduction in line-of-sight extinction to the infrared continuum, 
{\it and} intrinsic source brightening. 

The total line-of-sight extinction is harder to evaluate. 
A recent estimate by \cite{Carvalho2021} 
makes use of diffuse interstellar bands DIBs absorption features
to estimate $A_V= 2.7$ mag for $R_V=3.1$ and $A_V= 4.3$ for $R_V=5$.
This is absorption against the outburst optical continuum.
Here, we find that de-reddening the flux-calibrated optical outburst 
spectrum by $A_V$ values in the range 4-9 mag produces 
a spectral energy distribution at blue wavelengths that is consistent 
with unreddened FGK stellar templates  
with the higher $A_V$ values corresponding to the earlier templates).
Again, these estimates would be the current extinction 
to the source of the optical continuum.
An additional $A_V \approx 3$ mag is needed to describe 
the difference between the bright state and faint state infrared 
spectral slopes, meaning that the faint-state extinction was in the range $A_V= 6-12$ mag.  

This range of $A_V$ values is consistent with what can be inferred from the 
[\ion{Fe}{2}] lines in the faint-state (2003) infrared spectrum.
We use the formalism of \cite{Pecchioli2016} with the intrinsic line ratios of \cite{Bautista2015}
and reddening law of \cite{Fitzpatrick1999}. 
From the measured [\ion{Fe}{2}] 1.257, 1.321, and 1.644 $\mu$m line fluxes of
1.47, 0.47, and 2.13 $\times 10^{-14}$, respectively, we find values of $A_V$= 7.3 $\pm$ 0.1
for $R_V = 3.1$. We can not derive a value for the bright-state (2019) spectrum
since the continuum brightening renders the [\ion{Fe}{2}] lines immeasurable.

%

It is unclear how much of the total line-of-sight extinction should be 
attributed to 
the molecular cloud and how much to the circumstellar environment, but we
note that the long-wavelength portion of the pre-outburst spectrum 
(Figure~\ref{fig:H2imageGspec}) does not show evidence of the $3\ \mu$m
water ice feature, which is often associated with high-extinction,
high-density environments.   The lower among the above extinction values
may thus be most appropriate.

\subsection{Bright State Quasi-Periodicity}

\begin{figure}
\includegraphics[width=0.48\textwidth]{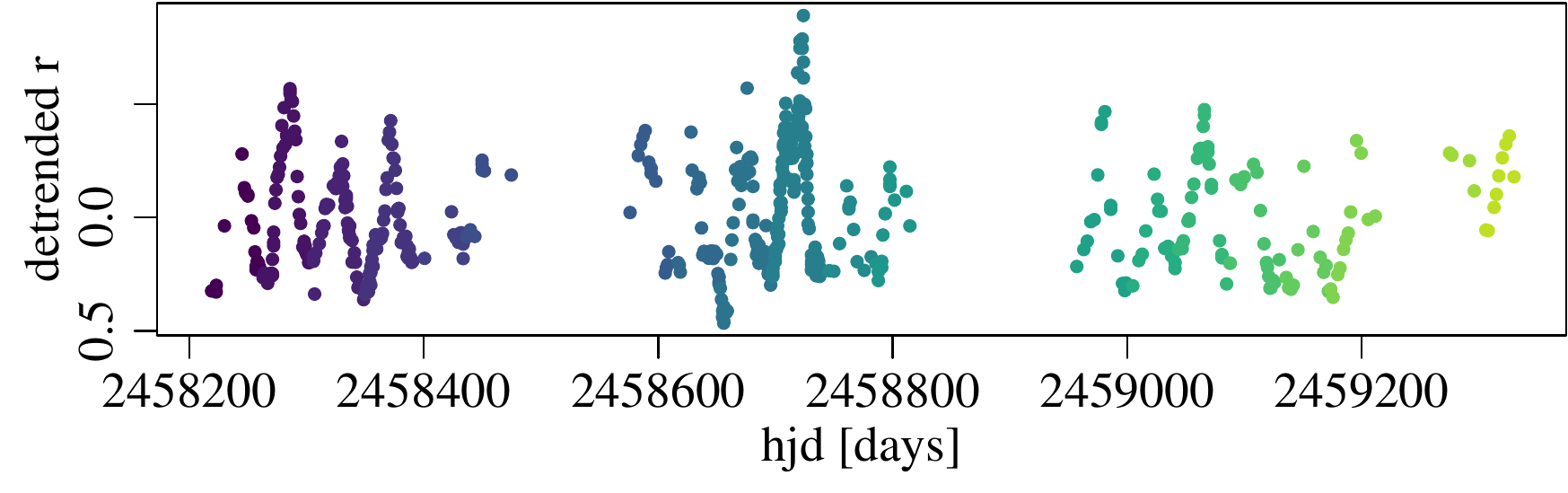}
\\
\\
\includegraphics[width=0.48\textwidth]{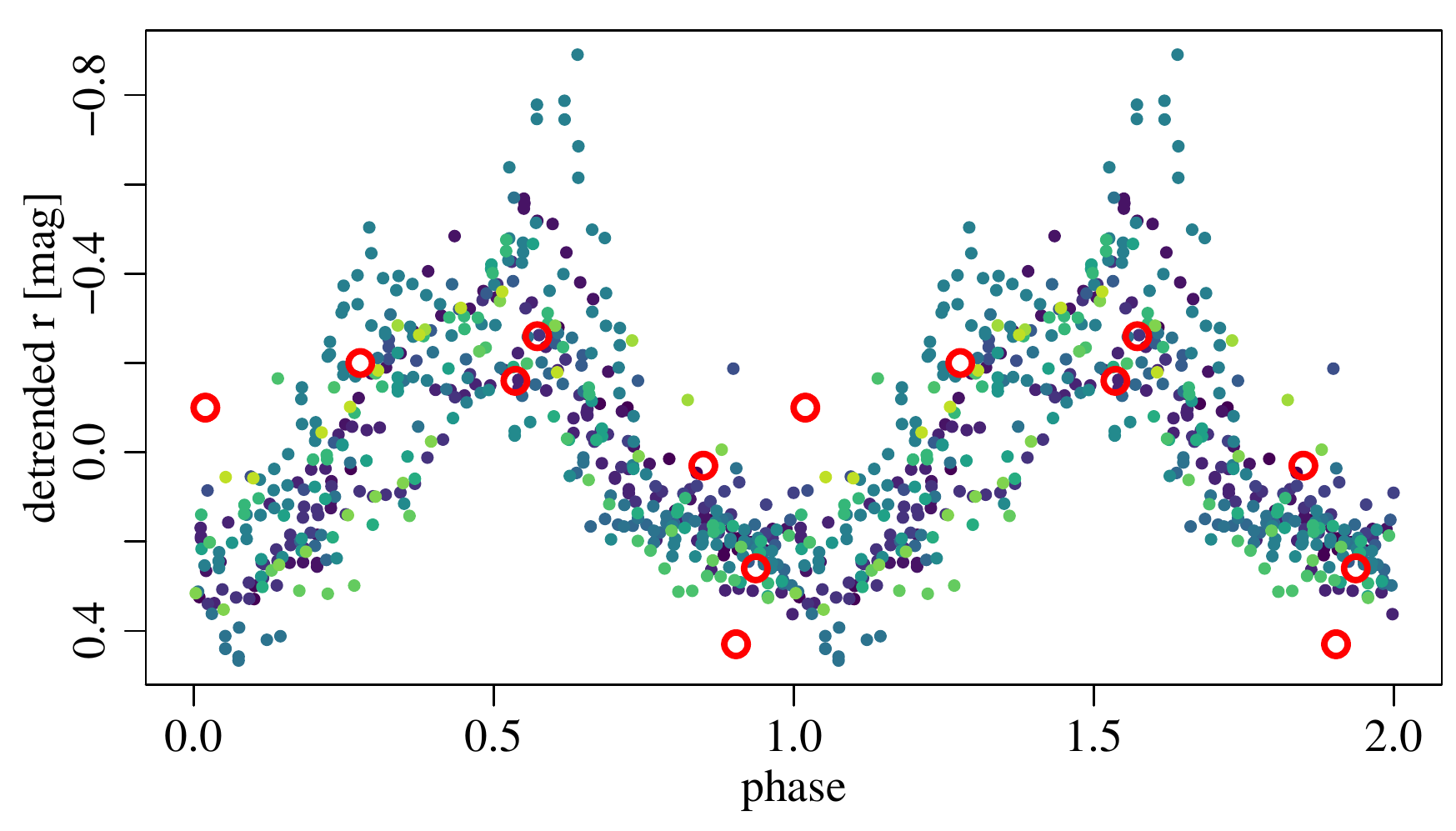}

\caption{
Top: ZTF $r$-band light curve with linear detrending performed separately for 
the 2018, 2019, and 2020 seasons.  Color scale shows a progression of 
hues from earlier to later cycles.
Bottom: Phased light curve for a period of 43.4 days, repeated twice so as
to more readily discern the full-phase shape.  Larger red points are the
seven data points from \cite{m2019}.
}
\label{fig:phased}
\end{figure}

To assess the periodicity of the photometric oscillations seen 
in Figure~\ref{fig:lczoomasas}, we employed a Lomb-Scargle period search algorithm
\citep[see e.g.][for a thorough discussion]{vdp2018}
as implemented in the LombScargleFast routine under
the gatspy.periodic python package.  The period was fit in flux units 
rather than in magnitudes.  Using the full $ZTF$ data set,
a single strong peak in the periodogram corresponds 
to a derived period of 43.4 days.  While the period has persisted for
three years, 
there is some evidence for a fluctuation over time in the mean source brightness,
on top of which the periodicity resides.  Including a fit for
this drift does not change the mean period or amplitude, but does reduce
the scatter as a function of phase.
The mean r-band magnitude was 13.0~mag during the 2018 season, 
which brightened to $r=12.6$~mag during the 2019 season, 
then faded to $r=13.3$ during the 2020 season.  In addition, within the seasons, 
we find that in 2018 there was a downward drift corresponding to 0.42 mag/yr, 
in 2019 an upward drift of -0.32 mag/yr, and in 2020 the source was fading
by 0.22 mag/yr. 
Correcting for these drifts and phasing the light curve results in the profile
shown in Figure \ref{fig:phased}.

Among individual cycles, the amplitude of the periodic signal ranges 
from $\sim$0.6 to $\sim$0.8 mag.  
A formal analysis using a smoothed version of the light curves
yields peak-to-trough amplitude of $0.66\pm 0.05$ mag for $r$ band 
and $0.68\pm 0.14$ mag for $g$ band.  The light curve peak-to-trough
excursion time is nearly 1/2 of the period, with the derived time between
light curve maximum and light curve minimum $0.48\pm 0.10$ times the
period for both $r$ band and $g$ band.  Notably, the clear periodicity 
in the $g$ and $r$ light curves comes without any change in color. 
We find an essentially constant value over time of $g-r = 2.8$ mag (AB) 
and no evidence for color periodicity in the source\footnote{The only
color periodicity appears to be a few percent signal associated with 
the $\sim$28 day cycle of the moon phase. This period is also seen 
as a low level brightness periodicity, though at a factor of ten
lower significance than the $\sim$43 day astrophysical peak.}. 

We note that the current periodicity can not be recovered in available photometry
from earlier epochs (Figure~\ref{fig:lc}).  Specifically,
neither the 2015-2017 data of \cite{m2019} nor the 1980's data of \cite{shevchenko1993}
show detectable periodicity.
However, we have checked whether the \cite{m2019} data can be phased
to the currently observed period. As shown in Figure~\ref{fig:phased},
the fluctuations seem significant and tend to display the same trends 
as the current data, but phase up with two of the seven points having
more scatter than the recent ZTF data.

\section{Spectroscopic Analysis of \lks}

\begin{figure}
\begin{center}
\includegraphics[width=0.9\textwidth,trim={0.75cm 0 0.75cm 0},clip]{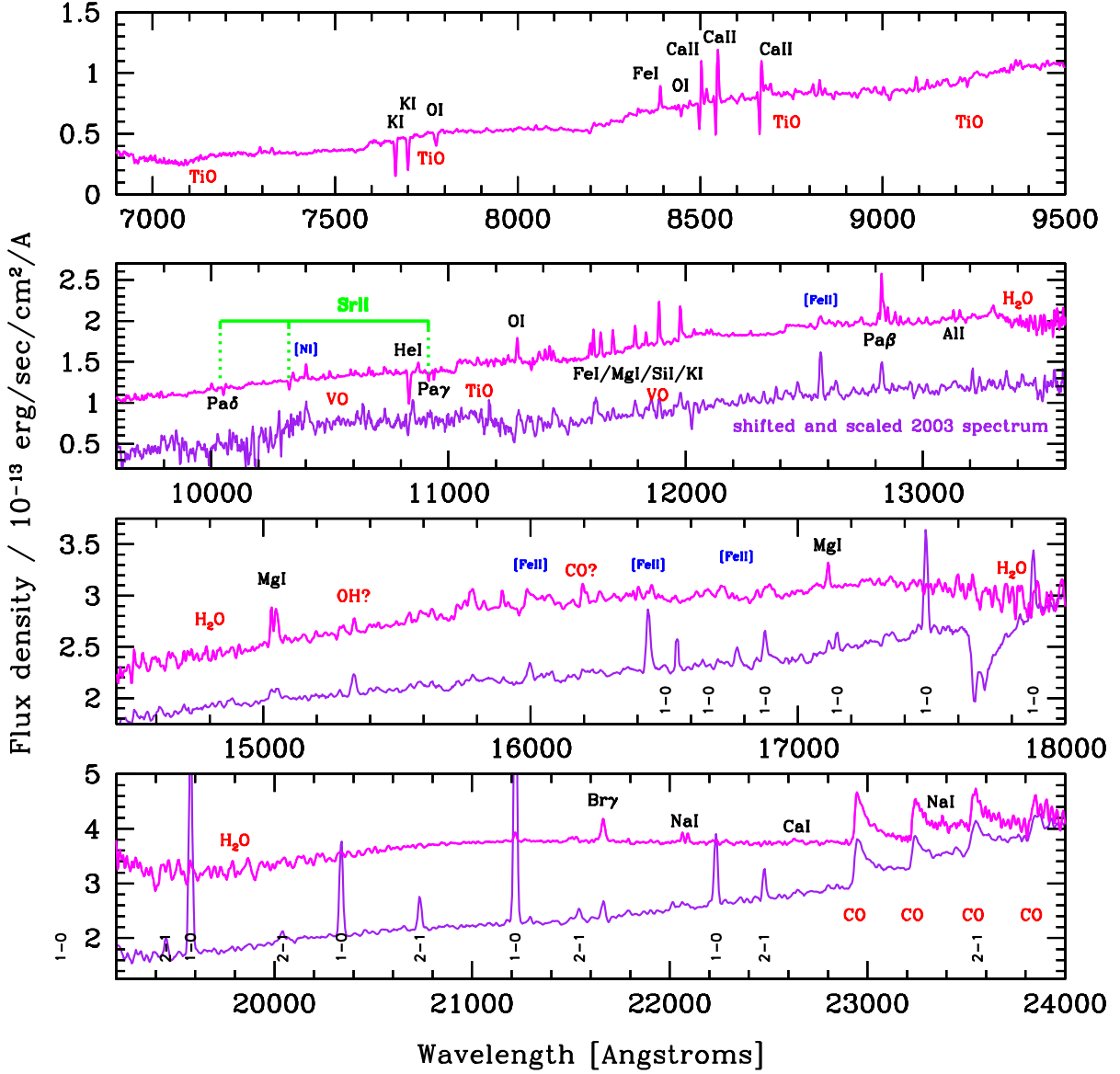}
\end{center}
\vskip-1.5truein
\caption{
IRTF/SpeX data comparing the 2019 outburst spectrum (upper, magenta) 
with a version of the 2003 faint-state spectrum (lower, purple) that has been
scaled (by a factor of 50) and shifted (individually in each panel). 
In the faint state, \lks\ exhibited relatively weak atomic emission features 
but strong molecular emission from H$_2$ (labeled as 2-1 or 1-0) and 
CO bands.  During the brightening from 2003 to 2019, 
the \lks\ infrared continuum has become bluer though it is still 
extremely red on an absolute scale.  The line-to-continuum ratio of the 
weak atomic emission has increased, while the line-to-continuum of the 
previously prominent H$_2$ lines has become weaker;  
the CO line-to-continuum appears relatively unchanged.
}
\label{fig:spexcompare}
\end{figure}

\begin{figure}
\begin{center}
\includegraphics[width=0.49\textwidth]{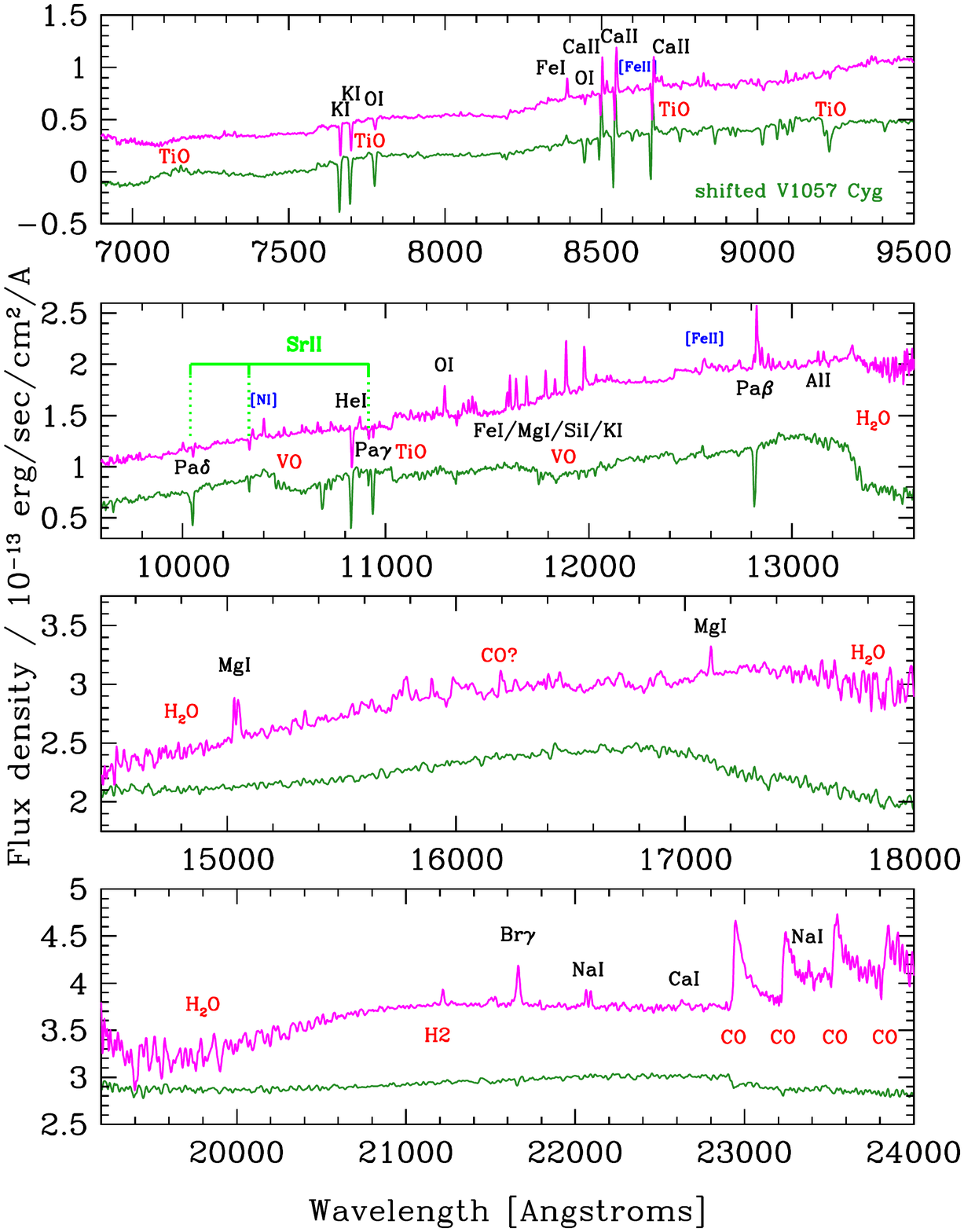}
\includegraphics[width=0.49\textwidth]{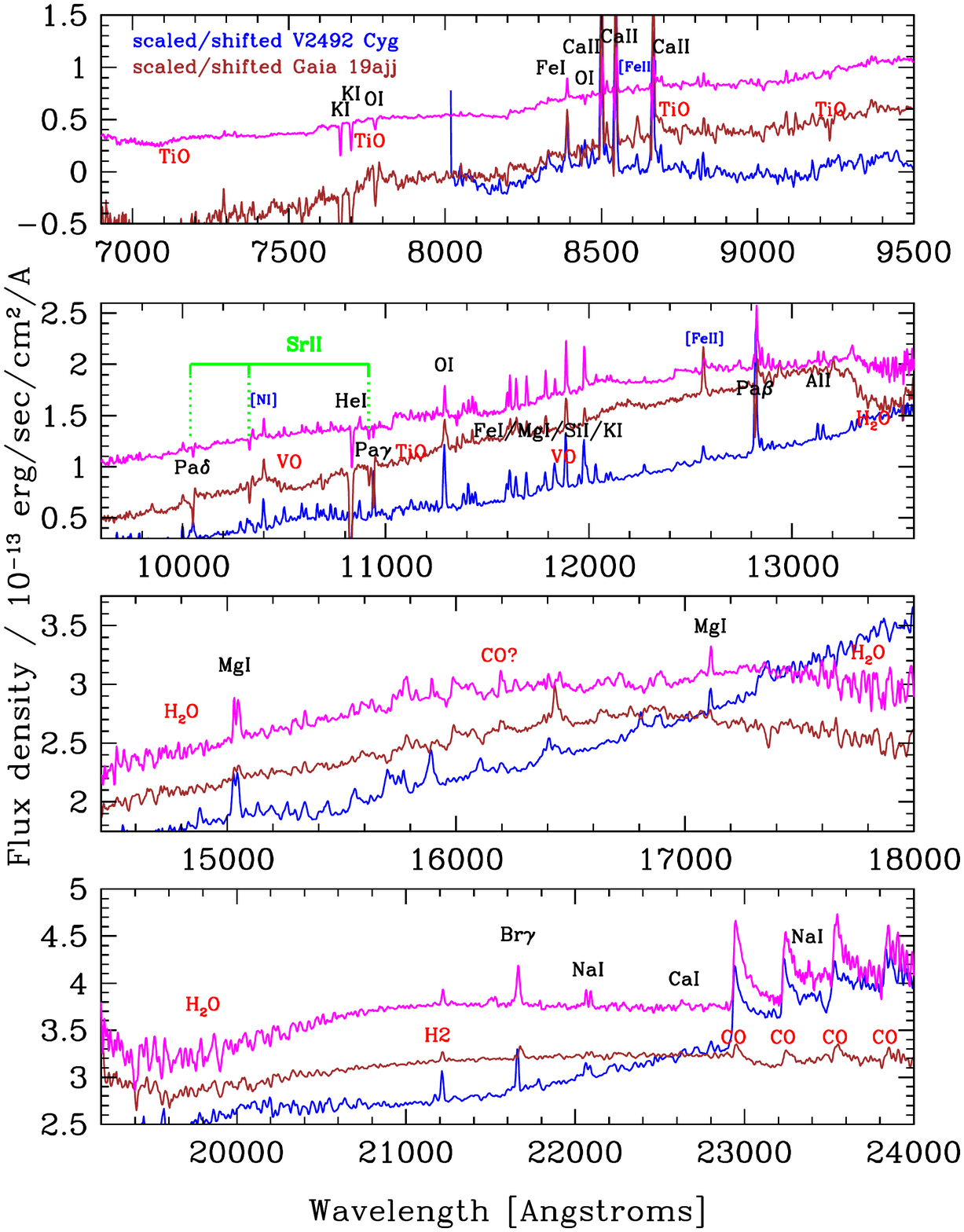}
\end{center}
\vskip-0.3truein
\caption{
Left panels:
Comparison of \lks\ (magenta) and V1057 Cyg (green) at infrared wavelengths. 
These sources are a reasonable match in terms of their absorption features
at blue optical wavelengths in low (Figure~\ref{fig:blueandtio})
and high (Figure~\ref{fig:hires}) spectral resolution data. 
This is also true at red optical wavelengths (top panel), 
including the TiO {\it emission} shared by both sources.
At $>9000$ \AA, however, 
the spectral match is less good, with \lks\ lacking 
the strong molecular H$_2$O, as well as the TiO, VO, and CO absorption 
that is seen in V1057 Cyg at these longer wavelengths.
The redward side of the J-band and the H-band does indicate some weak H$_2$O,
however.  And rather than CO absorption, \lks\ has strong CO emission, 
as well as other atomic emission lines that are more reminiscent 
of other types of young star outbursting sources.
Right panels:
The same \lks\ spectrum as in the left panels 
(magenta) now compared to V2492 Cyg (blue) and Gaia 19ajj (brown). 
\lks\ shares the weak H$_2$O absorption in the J-band and H-band 
with Gaia 19ajj, though the TiO and VO patterns differ slightly.
The CO emission of \lks\ is more similar to V2492 Cyg than Gaia 19ajj,
and the atomic emission line patterns better match this source as well.
The only atomic absorption in \lks\ is from \ion{Sr}{2}
(see also Figure~\ref{fig:nirspec}), 
which is also seen in both V1057 Cyg and Gaia 19ajj.
}
\label{fig:irspec}
\end{figure}

The subsections below present the salient details of the \lks\ spectral features.  
We begin by comparing the outburst infrared spectrum 
to an earlier pre-outburst spectrum, highlighting changes.
We then proceed to discuss the various spectral elements of the outburst spectrum
in detail.  There is evidence of a strong wind/outflow seen against the
optical continuum, but not of a shocked atomic gas component, 
which would manifest as e.g. strong forbidden lines formed in a jet.  
However, there are weak shocked gas emission signatures in molecular H$_2$ 
and atomic [\ion{Fe}{2}] in the infrared that have weakened relative 
to the continuum during the outburst.  The outburst also features
molecular TiO and CO emission and weak atomic metal line emission 
that appears mainly at wavelengths longer than $\sim$6000 \AA.
Finally, there is photospheric absorption 
that apparently arises from a low-gravity atmosphere having mixed temperature,
with both hot features (e.g. \ion{Fe}{2}) at blue wavelengths 
and cool features (e.g.  H$_2$O) at red wavelengths.
After the following detailed description of the outburst spectrum, we close the section
by describing the subtle differences among the three HIRES spectrum
taken at moderately different photometric phases.

\subsection{Spectral Changes between 2003 and 2019}

The spectral ratio analysis described above reveals that the H$_2$O absorption 
seen in Figures \ref{fig:spexcompare} and \ref{fig:irspec}
has been enhanced during the outburst. Another finding 
is that there is essentially no change in the CO line-to-continuum, 
with only continuum remaining in this wavelength region of the ratio spectrum.
Any change in TiO is hard to assess, as it is not possible to determine
whether or not the source had TiO before the long-duration outburst. 
Also, the prominence of narrow H$_2$ emission relative to the continuum 
has decreased in the bright state. 

Regarding the atomic emission, detailed examination shows that essentially all of 
the 1-2.4 $\mu$m lines now seen in emission in the bright state were also present 
in the faint state.  The atomic emission line pattern
is essentially the same, only weaker in the earlier spectrum.
The ratio spectrum shows no change in the line-to-continuum,
i.e. no signature of the emission lines in the ratio spectrum.  The implication is that 
the photometric brightening contributes equally to the continuum and the atomic line emission.  

Going back even further, 
the near-infrared spectra shown in \cite{aspin1994}, taken in 1991, 
compare very well to those illustrated in
Figure~\ref{fig:H2imageGspec}, taken in 2003.  
The earlier date closely follows the end of the \cite{shevchenko1993} 
time series data illustrated in Figure~\ref{fig:lc},
which is just before the deep fade\footnote{\cite{aspin1994} report
fading by 2 mag in the near-infrared
between October 1991 and November 1993.} 
from $\sim 15^m$ to $\sim 19^m$.   The 2003 date was firmly in the deep fade. 
Thus, the fading and accompanying reddening \citep{herbst1999}
do not seem to have resulted in any change in the source spectrum.
Conversely, the brightening from $\sim 19^m$ to the current $\sim 13^m$
has produced a change in the emission and absorption spectrum
in the $K$-band (Figure~\ref{fig:irspec}).

\subsection{Wind and Jet Outflow Lines}

\begin{figure}
\includegraphics[width=1.00\textwidth]{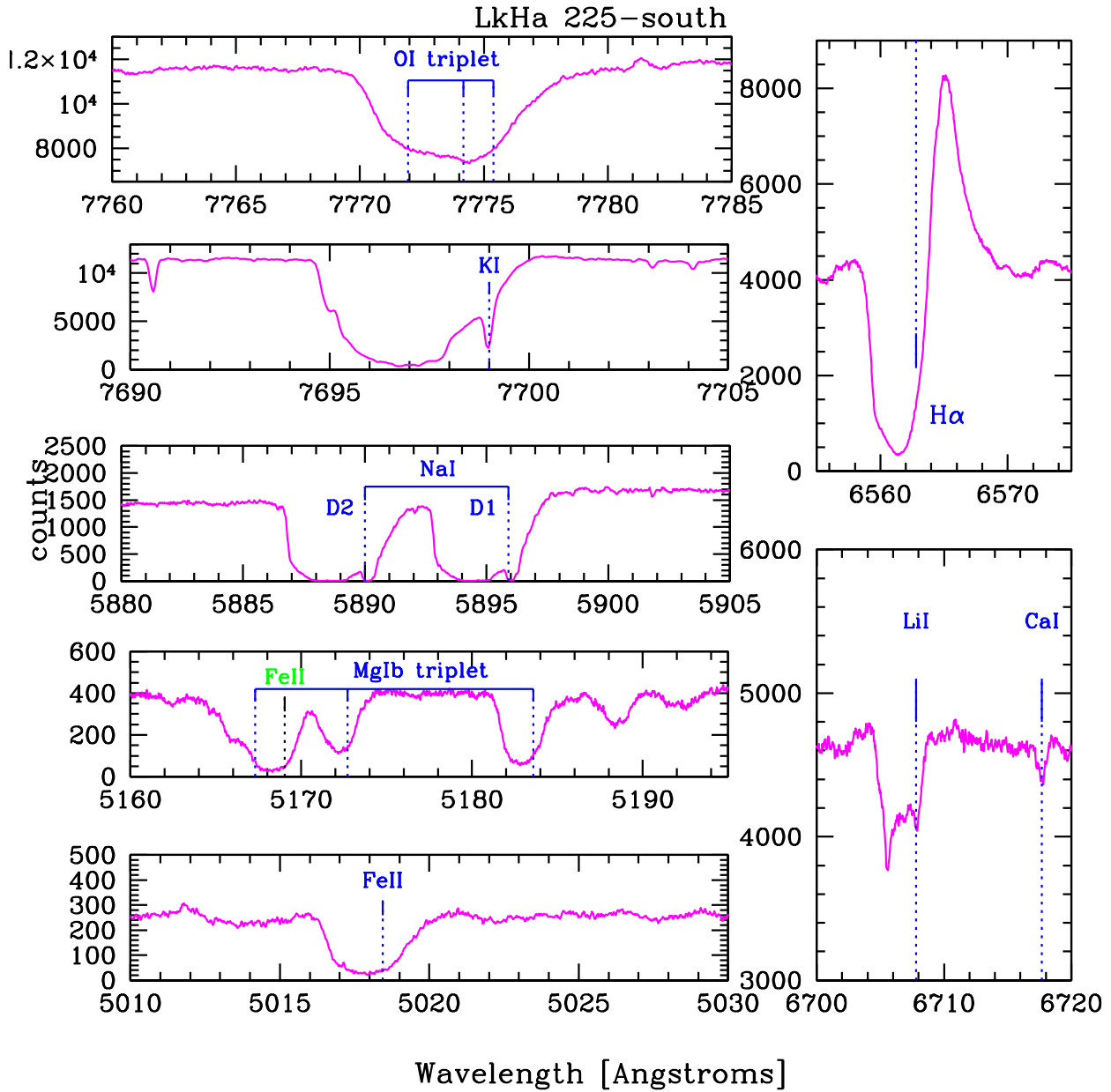}
\vskip-1truein
\caption{Portions of the August 2019 Keck/HIRES spectrum illustrating the strong wind of \lks\
via blueshifted absorption features.  
In the lower right panel, the narrower \ion{Ca}{1} absorption line 
validates the wavelength scale and thus our heliocentric correction 
to the stellar rest velocity.
The H$\alpha$ profile (upper right panel) has a clear P Cygni nature, 
with the absorption against the emission profile extending to 
about $-210$ \kms\ on the blue side,
though the emission wing continues to about $-325$ \kms. 
The red side emission peaks at $+85$ \kms\ and extends to +275 \kms.
Broad blueshifted absorption is also seen in \ion{Li}{1} (lower right panel),
and the \ion{K}{1} doublet, the \ion{O}{1} triplet,
the \ion{Na}{1} D doublet lines, 
the \ion{Mg}{1} b triplet lines, and among the wind-sensitive \ion{Fe}{2} lines. 
The typical terminal velocity in the metal lines is $-150$ \kms. 
}
\label{fig:wind}
\end{figure}

\begin{figure}
\includegraphics[width=0.70\textwidth]{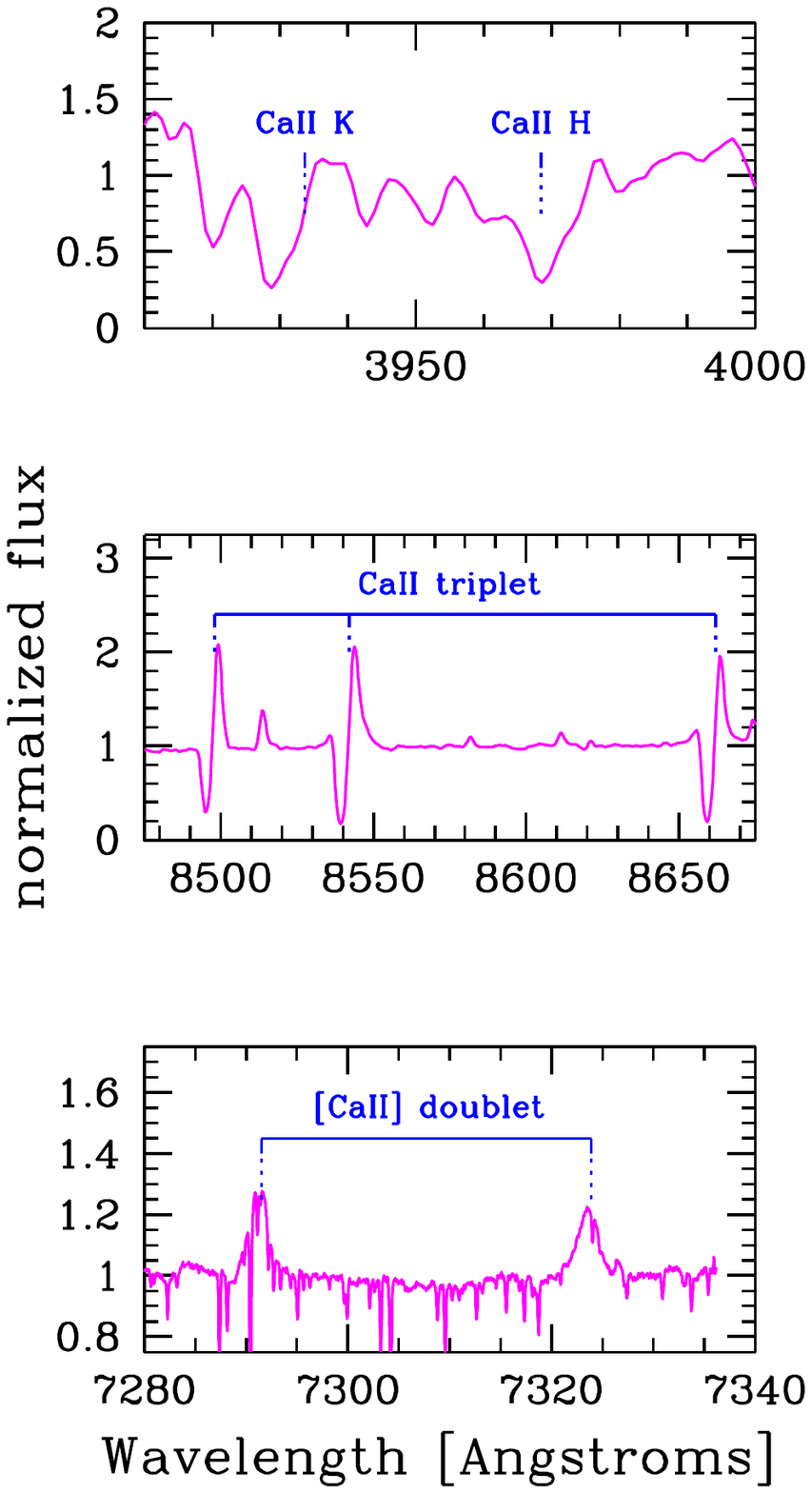}
\vskip-1truein
\caption{The optical ionized calcium lines in \lks.
Top and middle panels show the permitted transitions of \ion{Ca}{2}
as seen in the Palomar/DBSP spectrum, while bottom panel shows 
the forbidden transitions of [\ion{Ca}{2}] as seen in the Keck/HIRES spectra
(with uncorrected telluric absorption contamination still present 
in the data).}
\label{fig:ca}
\end{figure}

Seen against the very red continuum are a number of
indicators of rapidly outflowing gas from \lks.
Figure~\ref{fig:wind} illustrates 
deep, blueshifted absorption in the usual prominent optical 
wind lines, notably \ion{Mg}{1}b, \ion{Na}{1}D, 
\ion{K}{1} 7665 and 7699 \AA, and \ion{O}{1} triplet.
Wind signature is even seen in \ion{Li}{1} 6708 \AA.
H$\alpha$ along with the \ion{Ca}{2} triplet lines (Figure ~\ref{fig:ca})
exhibit P Cygni structure.

The \ion{Na}{1}D wind absorption is saturated, and the \ion{K}{1} nearly so.
We note explicitly that the wind evidenced in the red lines of
\ion{K}{1}, \ion{O}{1}, and \ion{Ca}{2} is forming against the continuum
of the molecular TiO/VO emission.  
At bluer wavelengths, absorption in lines of \ion{Fe}{2}, 
such as 4172, 4179, 4924, 5018, 5169, 5197, 5234, 5316 \AA\ 
is evident in our low resolution spectra 
(e.g. Figure~\ref{fig:blueandtio}) and revealed in our higher resolution
spectra (not shown) to also have broad blue asymmetry, indicative of a wind.
The typical terminal velocity of these metal lines is about $-150$ \kms.

The H$\alpha$ and \ion{Ca}{2} triplet lines are strong and broad.
However, the entire blueshifted portion of the emission profile 
is not in fact absorbed.  Instead, the absorption extends to about -210 \kms\
with an additional wing of unabsorbed blueshifted {\it emission} 
that extends to about -325 \kms\ (see Figures~\ref{fig:wind} and ~\ref{fig:ca}.  
At low dispersion, our H$\alpha$ profile appears very similar 
to the time series of such profiles that is shown in Figure 4 of \cite{m2019}.

\begin{figure}
\includegraphics[width=0.45\textwidth,angle=-90]{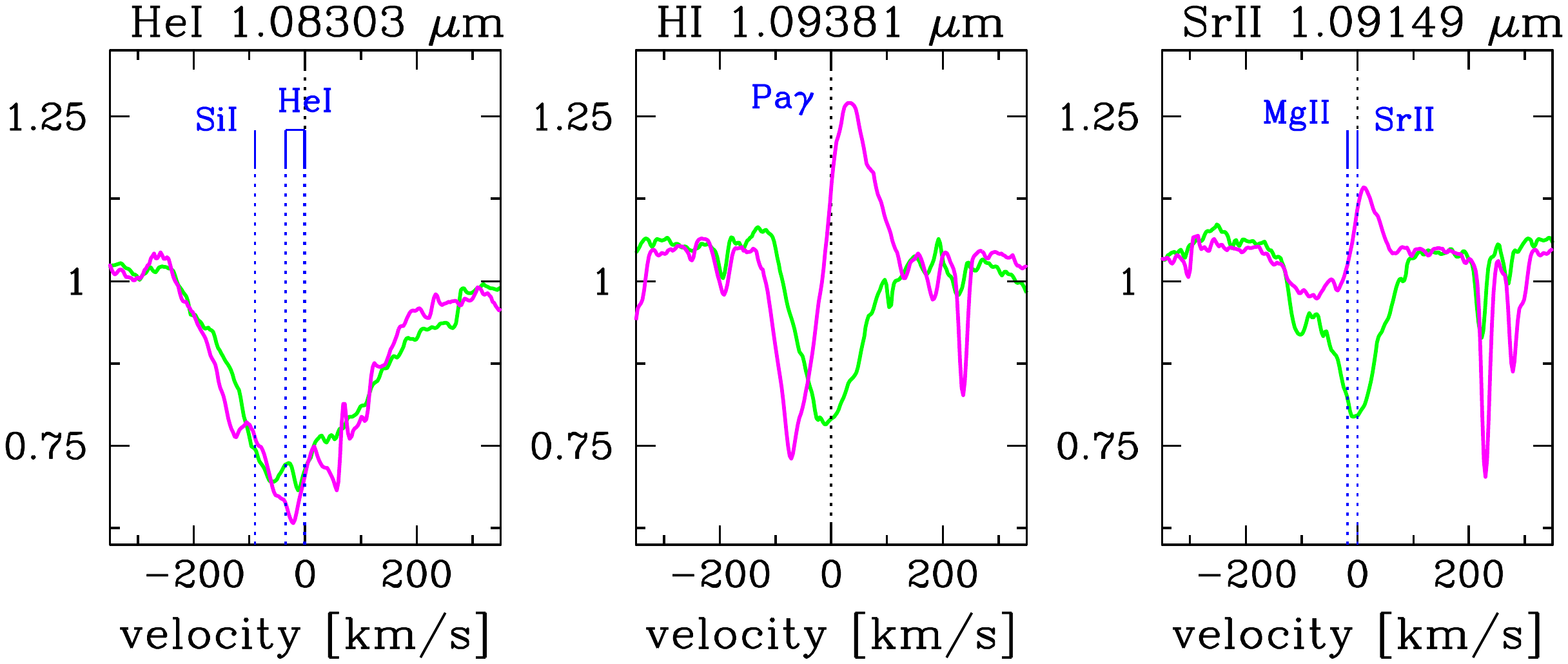}
\caption{
Keck/NIRSPEC line profiles for \lks\ in \ion{He}{1} 10830 \AA,
\ion{H}{1} Pa$\gamma$, and \ion{Sr}{2}. 
Magenta spectrum is from June 2019 while green spectrum is from September 2020.
In the \ion{He}{1} panel, the marked \ion{Si}{1} line at 10827 \AA\ 
is unlikely to be a contributor to the observed profile, 
given the absence of comparable-strength \ion{Si}{1} lines at 10844 and 10870 \AA.
Similarly in the \ion{Sr}{2} panel, the marked \ion{Mg}{2} line at 10914 \AA\ 
is unlikely to be a contributor to the observed profile, 
given the absence of a comparable-strength line at 10952 \AA.
The narrow absorption features in the Pa$\gamma$ and \ion{Sr}{2} panels are uncorrected telluric lines.  
While the \ion{He}{1} profile shows no change, the \ion{H}{1} line
has evolved from a P Cygni profile to pure absorption.
The \ion{Sr}{2} profile has evolved similarly. 
These line profiles can be compared to those exhibited by
the cooler optical wind lines of Figure~\ref{fig:wind}. 
}
\label{fig:nirspec}
\end{figure}

In the infrared, Figure~\ref{fig:irspec} indicates Br$\gamma$ and Pa$\beta$
strongly in emission, but the resolution is too low to discern the line
kinematics.  The Pa$\gamma$ line is ambiguously present in Figure~\ref{fig:irspec}, 
but Figure~\ref{fig:nirspec} shows this line at higher dispersion. 
It has a clear P Cygni profile indicating wind signature in our 2019 spectrum,
but the 2020 spectrum shows the line in pure absorption, centered at zero velocity.

Figure~\ref{fig:nirspec} also shows the \ion{He}{1} 10830 \AA\ triplet profile,
which has a depth of about 70\% of the continuum and width $\pm 200$ \kms.
The profile is fairly symmetric around the weighted line center of the triplet.
Furthermore, the absorption is broader than that exhibited in the nearby hydrogen or metal lines. 
There is no obvious P Cygni structure or blueshifted asymmetry 
in \ion{He}{1} 10830. In other young star outbursters \ion{He}{1} 10830
line depths can reach to only $\sim$5\% of the continuum in some cases.
One object with a \ion{He}{1} 10830 profile that is quite similar to that 
of \lks\ is FU Ori itself, with similar centroid, width, and depth. 

Notably, there is no change in the \ion{He}{1} 10830 \AA\ profile between
our two spectra taken approximately 1.2 years apart.
As stated in the Introduction, an early spectrum of \cite{andrillat1976} also 
showed strong absorption in the \ion{He}{1} 10830 \AA\ line. 
There is indication of such in the 2003 IRTF/SpeX spectrum as well 
(see Figure \ref{fig:spexcompare}), possibly with a P Cygni type profile,
which is not present in the current outburst state.

Regarding forbidden line emission, we do not see a strong signature 
in the [\ion{O}{1}] or [\ion{S}{2}] lines that are reported as
detected by \cite{m2019}. However, these authors acknowledge that bad
sky subtraction could be an issue in their data, and could cause 
a false positive.  The source is not significantly different in brightness
now compared to the \cite{m2019} observations, so continuum brightening
is not the explanation for the lack of detection of these lines in our data.
We do find, however, unambiguous [\ion{Ca}{2}] doublet
emission at 7291,7324 \AA\ (Figure~\ref{fig:ca}).  
This doublet is rarely seen in normally accreting young sources, 
though was seen in the outbursting  
case of PTF~14jg \citep{hillenbrand2019a}. 
The [\ion{Ca}{2}] doublet is the only strongly detected optical forbidden line 
in the outburst state of \lks.  

\subsection{Molecular Emission}

Molecular emission in the form of H$_2$ is associated with shocked gas 
in near-circumstellar environments. This molecule is only weakly seen 
in the $K$-band region of \lks\ in its outburst state. 
Comparing the recent outburst spectrum to the earlier 
2003 pre-outburst or early outburst spectrum, the H$_2$ line-to-continuum
is weaker in the outburst spectrum, as discussed above.

More unusual, in terms of molecular features, is that in outburst 
\lks\ exhibits the rare phenomenon of having TiO bands in emission. 
This is seen from the red optical out to the 11300 \AA\ band
(Figures~\ref{fig:blueandtio} and ~\ref{fig:irspec}).  There is 
a hint that the 11300 \AA\ emission was also present in the pre-outburst state
(Figure~\ref{fig:spexcompare}).
Only a few young stars have been documented with TiO in emission, 
specifically PTF 10nvg / V2492 Cyg \citep{covey2011},  
IRAS~04369+2539 and IRAS~05451+0037 \citep{hillenbrand2012}, 
VV CrA \citep{herczeg2014}, 
iPTF15afq \citep[][a.k.a. Gaia19fct]{miller2015}, and PGIRN 20dwf \citep[][a.k.a. Gaia20eae]{hankins2020}. 
Figure~\ref{fig:irspec} demonstrates V1057 Cyg to be in this category as 
well; the spectrum is the same as that shown in \cite{connelley2018}, 
though the TiO emission aspect was not commented upon in that paper.  

Typically, TiO and VO bands are seen in absorption, 
and signify the presence of warm molecular gas ($\sim 1500-4000$ K)
in a cool stellar atmosphere.  These molecules define 
the onset of the type M spectral class, for example, with increasing 
absorption seen in the TiO/VO species towards later types, 
transitioning to CO and H$_2$O molecules at even cooler M-type photospheres.
All four of these molecules are observed in absorption
from the disk atmospheres associated with FU Ori stars \citep{connelley2018}.

In \lks, we see the optical and near-infrared TiO/VO bands,
and the infrared CO lines, all in emission, 
but the H$_2$O molecule is in absorption.

\subsection{Atomic Emission}

There is sparse, weak narrow atomic emission in the outburst spectrum of \lks. 
The lines are at the rest velocity, and overall the narrow emission spectrum 
is quite similar to that of V2492 Cyg, suggesting a similar temperature,
though weaker in terms of line strength.

In the optical, the emission is seen mainly at the redder wavelengths 
of Figure~\ref{fig:blueandtio}, with little emission 
detected shortward of $\sim$6000 \AA, even at high spectral resolution.  
The optical lines are mainly \ion{Fe}{1} with some \ion{Ni}{1} and \ion{Ti}{1}. 
There is an apparent trend in emission vs absorption with 
line excitation energy. Many lower excitation \ion{Fe}{1} and \ion{Ni}{1} 
lines are weakly in emission, but the higher excitation lines in these same
neutral species are seen mainly in absorption.

At infrared wavelengths, as depicted in \ref{fig:spexcompare},
there is a slightly stronger, well-populated atomic emission line forest 
in the J-band.  There is also weak \ion{Mg}{1} in the H-band, 
and weak \ion{Na}{1} and \ion{Ca}{1} in the K-band.

\cite{m2019} reported that many (optical) lines of \ion{Fe}{2} 
are in emission, which we do not see.  Our spectra were taken 
approximately 4 years later and show these same \ion{Fe}{2} lines 
ubiquitously and unambiguously in absorption. 
This suggests cooling spectral evolution after the source reached its peak 
brightness. However, the low excitation \ion{Fe}{1} lines remain in emission.  
The situation in \lks\ seems somewhat similar to the one in PTF~14jg, 
which was categorized by \cite{hillenbrand2019a}
as an unusually hot accretion outburst.  That source exhibited
strong absorption in \ion{Fe}{2}, as well as in other even hotter ionized metals, 
early on in its outburst.  PTF~14jg also showed \ion{Fe}{1} 
and other neutral metal emission lines at the same early times, but these
lines faded over several years and became absorption features.

\subsection{The Pure Absorption Spectrum}

\begin{figure}
\includegraphics[width=0.60\textwidth,angle=-90,trim={0 0 0 0.75cm},clip]{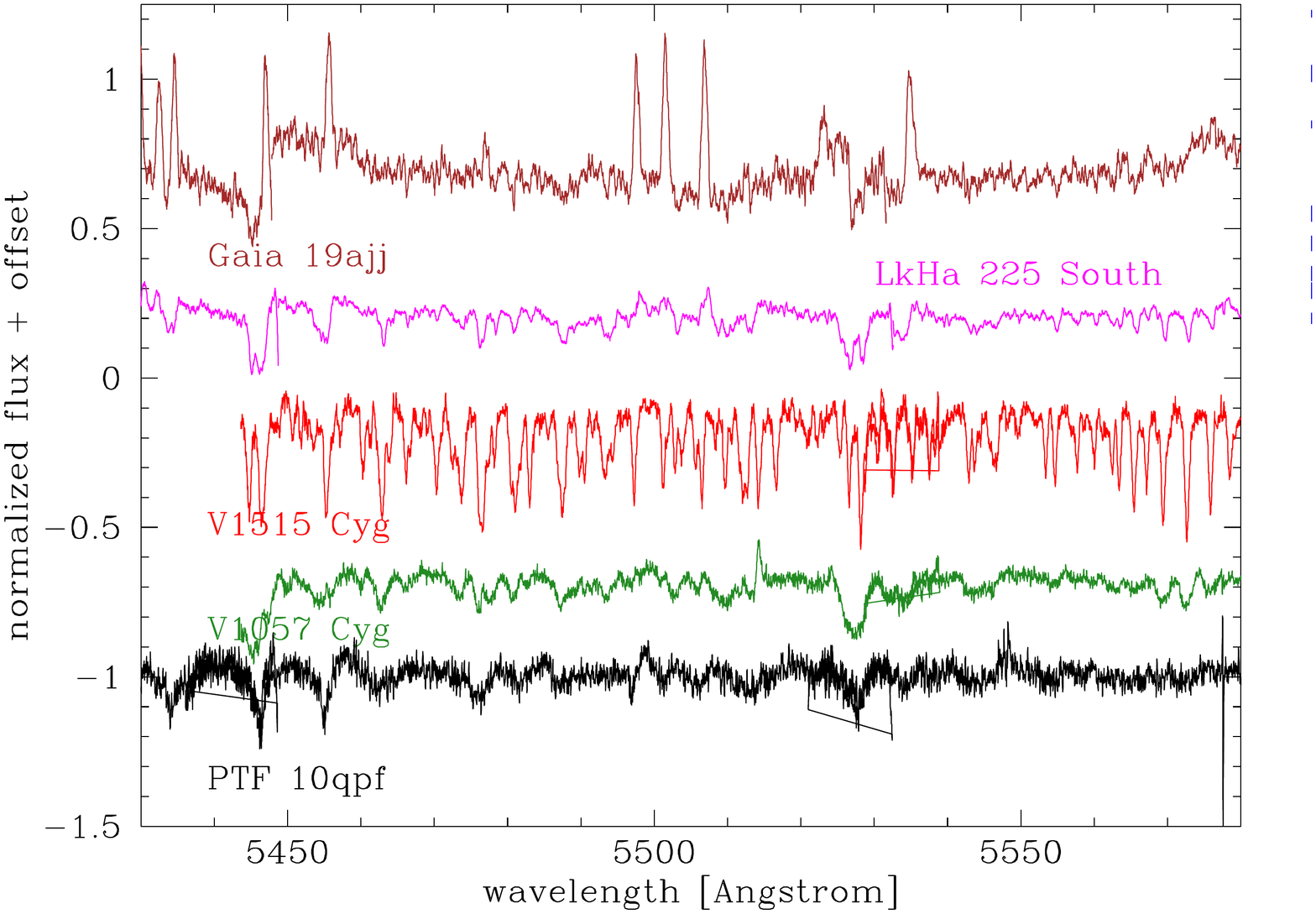}
\includegraphics[width=0.60\textwidth,angle=-90,trim={0 0 0 0.75cm},clip]{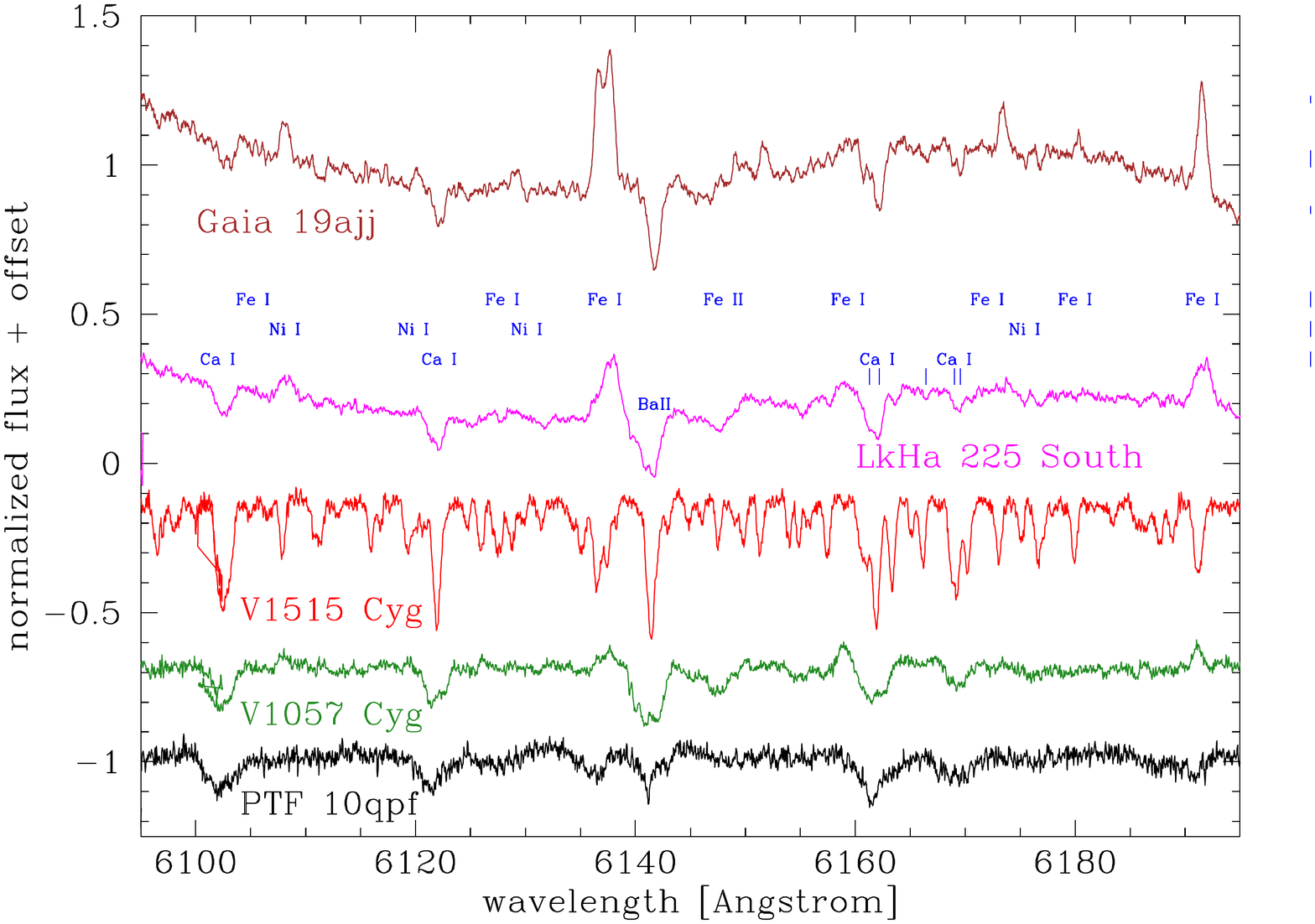}
\caption{
Portions of the August 2019 HIRES spectrum highlighting the spectral similarity
of \lks\ (magenta) to other outbursting sources.
The absorption component in \lks\ bears resemblance at shorter wavelengths
(e.g. top panel)
to that of the FU Ori stars V1515 Cyg, V1057 Cyg, and V2493 Cyg (PTF 10qpf, HBC 722). 
At longer wavelengths (e.g. bottom panel) the absorption component is less 
obvious, but still present in the stronger lines. 
Note in particular the presence of \ion{Ba}{2}, which is 
atypical in non-outbursting T Tauri or Herbig Ae/Be stars, 
but generally is seen in FU Ori objects.  
In addition to the weakening absorption component towards longer wavelengths,
there is increasingly weak emission, which can be appreciated by
comparison with another recent outbursting source, Gaia 19ajj.  
Like \lks, this source has a mixed spectrum of absorption and emission lines
in neutral metals, with overall stronger emission and weaker absorption 
relative to \lks. 
}
\label{fig:hires}
\end{figure}

The \lks\ absorption spectrum is peculiar, and not readily interpreted
as either a normal single star, or a binary star system.  Instead, we suggest
a more complicated composite spectrum.

Atomic absorption lines become increasingly prominent towards shorter wavelengths -- 
down to the $<4000$ \AA\ limit of our data.
The photosphere of \lks\ is thus apparently not heavily veiled in the blue.
Veiling, or accretion continuum emission that occupies a low filling factor, 
would weaken the absorption lines.  It arises under situations 
where the accretion rate is modest.

For higher accretion rates, it is possible to produce a predominantly absorbing
spectrum that originates in the hot inner regions of an accretion disk. 
In this scenario, the bluer wavelength absorption lines we observe could
originate in such a disk, or a wind emanating from this same region.
A relatively cooler emission-line component would fill in the low excitation 
absorption lines that would otherwise be present, turning some into emission.
We present detailed inferences from the observed absorption spectrum 
in the sub-sections below.

In addition to the absorption lines that originate in the 
(possibly extended) \lks\ photosphere, 
we also note the presence in our high resolution optical spectra DIBs. 
\cite{m2019} also mentioned the presence of DIBs features in \lks.  
Specifically, we see the 6614, 6379, 5850 (weak), 5797, and 5780 \AA\ DIBs.
As mentioned above, these can be used to estimate the extinction
to the optical continuum source, as recently done for this source (and others)
by \cite{Carvalho2021}.

\subsubsection{Temperature}

The low resolution Palomar/DBSP spectrum 
(Figure~\ref{fig:blueandtio}) is clearly composite at blue wavelengths.
Readily visible are \ion{Fe}{1}, \ion{Mn}{1}, and \ion{Ca}{1}
amidst a forest of other absorption lines, including strong \ion{Fe}{2}
(a hotter line). 
The high dispersion spectrum (e.g. Figure~\ref{fig:hires}) 
allows us to identify additional lines, spectrally resolving them.
Many intermediate excitation neutral lines are seen in absorption. Among them
are the \ion{Fe}{1} 7569 \AA\ (4.3 eV) and \ion{Ni}{1} 7573 \AA\ (3.8 eV)
featured by \cite{petrov2008}.  Also present are various
\ion{Ca}{1} (2.5-3 eV), \ion{Mg}{1} (5-6 eV) and \ion{Fe}{2} (4-6 eV) lines. 
The redder range of the spectrum exhibits several even higher excitation lines 
like \ion{Ca}{2} 8912 and 8927 \AA\ (7.1 eV).
Perhaps bounding the temperature on the high side, the hot
\ion{Si}{2} 6347 and 6371 \AA\ (8.1 eV) lines that have been seen in objects
like PTF~14jg are not present in \lks.

To attempt to discern the approximate temperature of the \lks\ 
outburst, we compared the de-reddened version of the low resolution 
optical spectrum to single-temperature standard stars.  
Based on the strongest metal lines, a best-estimate spectral type 
across the blue optical is somewhere between a late F and an early K star. 
However, \lks\ is a poor match to any single-temperature template spectrum 
in this spectral range. 
The \lks\ spectrum additionally has a pattern of lines
that are matched only in mid- and late-A type stars
(appearing much stronger than in the standards) and other lines that
are matched only in mid- and late-K type stars.  We thus do not report
a spectral type for the outburst spectrum beyond ``mixed".

\begin{figure}
\includegraphics[width=0.70\textwidth,angle=-90]{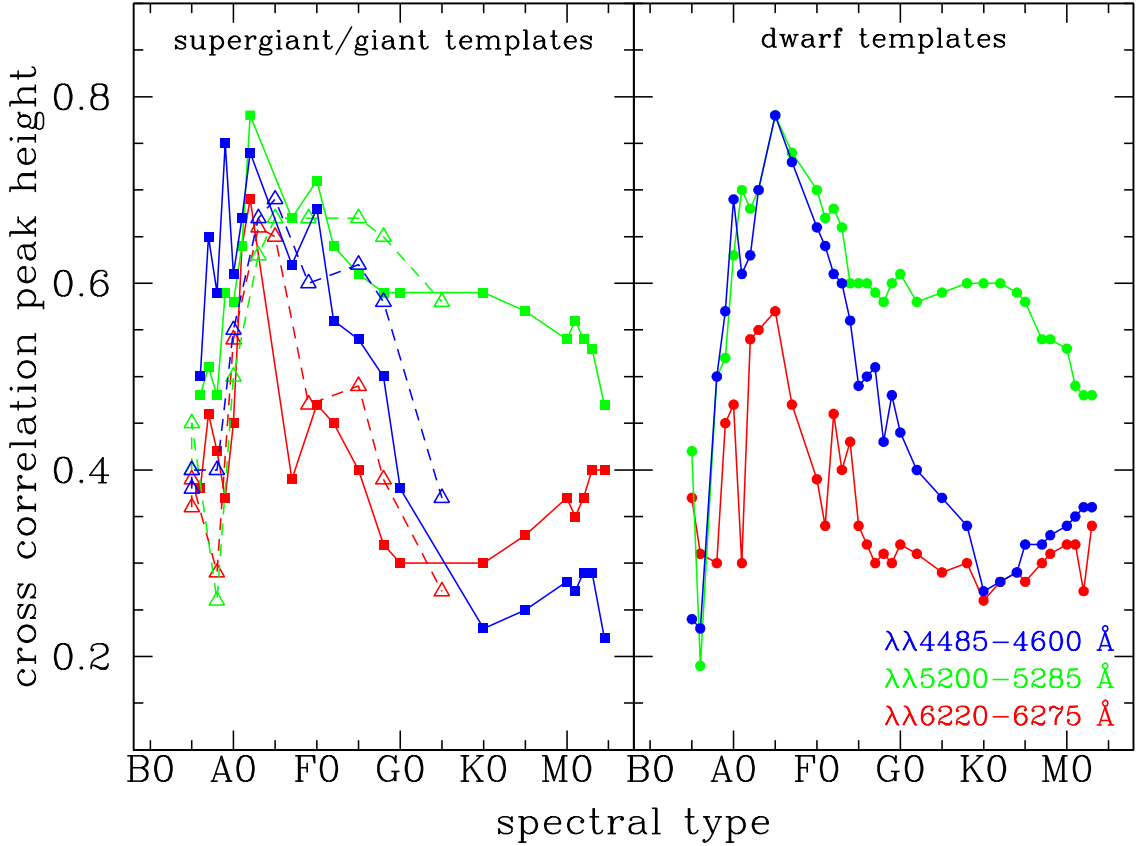}
\caption{
Resulting peak heights from a cross correlation of \lks\ Keck/HIRES spectrum
with a grid of spectral templates from the ELODIE spectral library,
plotted as a function of spectral type. Right panel shows dwarf (circles) templates
and left panel shows giant (squares) and supergiant (triangles) templates.
Blue points are from the
4485-4600 \AA\ region, green 5200-5285 \AA, and red 6220-6275 \AA.
For all wavelength ranges, the best correlations are for stars in the A3-A6
spectral type range, with no clear surface gravity preference.
The 5200 \AA\ spectral range has a broad plateau of good correlation
strength that extends throughout the FG and early K spectral types.
}
\label{fig:templates}
\end{figure}

To assess the spectral type at high dispersion, we perform a cross-correlation
analysis of the Keck/HIRES spectrum of \lks\ and a grid of stellar templates 
selected from the ELODIE spectral library \citep{prugniel2007}.
Figure~\ref{fig:templates} shows the results of this analysis 
over three different wavelength ranges, chosen to exclude spectral regions 
with strongly wind-affected species like those discussed in \S6.2 above.  
Based on this analysis, the best spectral type for the photosphere 
of \lks\ appears to be A3-A6 V.
Our quantitative finding can be confirmed by visual examination of the spectra. 
The pattern of metal absorption lines is well-matched
to the mid-A spectral type range, with similar strength lines 
that would be expected in FG spectral type comparison stars not as 
strongly present.  

A further conclusion from the cross correlation
analysis is that there seems to be little evidence for a systematically
changing spectral type with wavelength. 
Although we found evidence in the low dispersion spectroscopy for
a composite spectral type, no clear systematic behavior along these lines 
is obvious in the optical high dispersion data.
However, Figure \ref{fig:templates} does show that
the e.g. 5200 \AA\ spectral range has a broad plateau of good correlation
strength that extends throughout the FG and early K spectral types.
Thus the composite or mixed spectral type conclusion seems robust.

In contrast to the results from single temperature comparisons, 
the match of the \lks\ spectrum at blue wavelengths
to the FU Ori stars shown in Figure~\ref{fig:blueandtio}, is quite good. 
The spectrum shows essentially the same detailed absorption pattern 
exhibited by V2493 Cyg (PTF 10qpf; HBC 722) and V1057 Cyg. 
The same is true at high dispersion, as shown in  Figure~\ref{fig:hires}
which includes the two objects mentioned above as well as V1515 Cyg.
Overall, \lks\ appears intermediate between V1515 Cyg and V1057 Cyg
in its absorption spectrum, with line widths that are more like the former, 
and line depths that are more like the latter.  

In the infrared, besides the H$_2$O molecular absorption discussed above,
there is little basis on which to estimate a spectral type.  
The only narrow absorption that is plausibly identified can be associated 
with the \ion{Sr}{2} ion, discussed below as a gravity indicator.  
Neither hot features, which would be consistent with the early spectral type 
inferred above from the optical spectrum, nor cool features, such as those
which would be consistent with the weak H$_2$O absorption, are present.  
The lack of atomic absorption prevents us from quoting an infrared spectral type. 
Unlike the situation in the blue optical, the infrared spectrum of
\lks\ does not appear to be a good match to FU Ori stars (Figure~\ref{fig:irspec}).

\subsubsection{Gravity Indicators}

\lks\ has strong \ion{Sr}{2} lines.  In the optical, these occur at 
4077 and 4216 \AA\ (Figure~\ref{fig:blueandtio}),
and in FG stars they indicate low gravity when strong relative to the blueward
\ion{Fe}{1} 4046, 4063, and 4071 \AA\ or redward \ion{Fe}{1} 4271 \AA\ lines.
These optical \ion{Sr}{2} lines are from multiplet 1. The multiplet 2 
infrared \ion{Sr}{2} lines at 10328 and 10916 \AA\ (Figures~\ref{fig:irspec},~\ref{fig:nirspec}) 
are also present, and also indicative of low gravity.  They are prominent in A--M 
supergiants, moderate in F--M giants, and weak in F--early-K dwarfs \citep{sharon2010}.

\lks\ also shows \ion{Ba}{2} lines.  Figure~\ref{fig:hires} illustrates
the 6142 \AA\ line of \ion{Ba}{2} in \lks. Both this line and 
the 6497 \AA\ line (not shown) are near-ubiquitously present
in FU Ori type objects, and a signature of a low-gravity atmosphere. 


\subsubsection{Radial Velocity}

We derive a radial velocity for \lks\ from the first Keck/HIRES spectrum 
of $-13.5\pm 0.1$ \kms. This is determined 
using telluric lines as the zero point reference, 
as described in \cite{chubak2012}, and the offset from zero velocity of
the absorption spectrum. 
We note that our value is similar to the velocities quoted by \cite{m2019} 
based on measurements of \ion{Fe}{1} and \ion{Fe}{2} {\it emission lines} at
lower spectral resolution.

The three Keck/HIRES spectra span a nearly 6 month time period and show
no detectable shift in radial velocity, with an upper limit of a few \kms.

\subsection{Spectral Changes During 2019}

There are perceptible differences in absorption/emission line 
strength and morphology among our three Keck/HIRES spectra,
taken over about 6 months.  The DIBs that are present show no changes, 
however, giving us confidence in our findings below regarding the 
small photospheric and wind line differences. 

Adopting a period of 43.40 days, as derived above, the three HIRES observations
taken at HJD = 2458713.8, 2458816.8, and 2458851.8 
span three cadences (see Figure~\ref{fig:lczoomasas}). They 
correspond to phases of 0.43, 0.81, and 0.61, respectively, 
where phase 0 would be minimum light and phase 0.5 would be maximum light 
over the $\sim 0.7$ mag brightness swing.
There is no obvious correlation in the limited data set between spectroscopic
behavior occurring closer to phase 0.5, vs that closer to phase 0.0.

The overall trend with time, however,
is that the narrow emission lines, which are mainly \ion{Fe}{1}, 
became somewhat weaker.  At the same time, absorption lines such as from
\ion{Fe}{1}, \ion{Fe}{2}, and especially \ion{Ca}{1} became stronger/deeper.
There are also morphology changes in some of the wind lines,
with blueshifted absorption components that increased in strength
and redshifted emission components that weakened in strength
and became narrower.  These changes are at the $\sim$10\% level in weaker lines,
but $>$25\% in strong emission/absorption features.

Although only [\ion{Ca}{2}] was apparent in our first HIRES spectrum, 
other weaker forbidden lines appeared later.
Specifically, weak [\ion{O}{1}] 6300 \AA\ and 
very weak [\ion{Fe}{2}] 7155 \AA\ are seen in our second and third
HIRES spectra. In the infrared, the SPeX spectrum 
(Figure~\ref{fig:irspec}) taken between the first and second HIRES epochs
shows [\ion{N}{1}] at 10400 \AA, 
as well as several very weak infrared [\ion{Fe}{1}] 
and perhaps still the [\ion{Fe}{2}] 1.25 and 1.644 $\mu$m features.
All of these forbidden lines, in addition to others,
were seen in Gaia 19ajj \citep{hillenbrand2019b}.

\section{Spectral Energy Distribution Modelling}

We model the 0.4-2.4 $\mu$m spectral energy distribution of \lks\ using 
a Keplerian disk model.  A full description of the model and procedure 
is given in \cite{rodriguez2022}, but we briefly summarize here.
We adopt a modified Shakura-Sunyaev temperature profile 
\citep{kenyon1988,zhu2007}, with each annulus radiating as an area-weighted spectrum given by a NextGen\footnote{BT-NextGen (AGSS 2009) available at \url{http://svo2.cab.inta-csic.es/theory/newov2/}} atmosphere \cite{Hauschildt1999} 
at the appropriate temperature. The model parameters are as follows. We set the source distance as fixed at $d=920$ pc, following \cite{m2019}, effectively the same as the $916\pm26$ pc argued above
based on the parallax of the nearby BD+40$^\circ$ 4124.
We assume an inclination i = 45$^\circ$.  This value is not well-justified,
but we were motivated to choose an inclination lower than the i = 60$^\circ$
suggested by a random distribution in sin$i$, based on the evidence presented
in \citep[e.g.][]{matthews2007} for an outflow orientation close to pole-on.
The outer radius of the disk is fixed at 100 $R_\odot$ but varying this parameter has no affect on the overall SED at the wavelengths of interest 
in this study. The mass of the central star, $M_*$, radius of the central star, $R_*$, accretion rate $\dot M$, and visual extinction $A_V$, however, are left as independent free parameters.
We also explore joint fitting of $M_* \dot M$, since that exercise produces corner plots that are less directly reflective of the prior assumptions.


\begin{figure}
\includegraphics[width=0.53\textwidth]{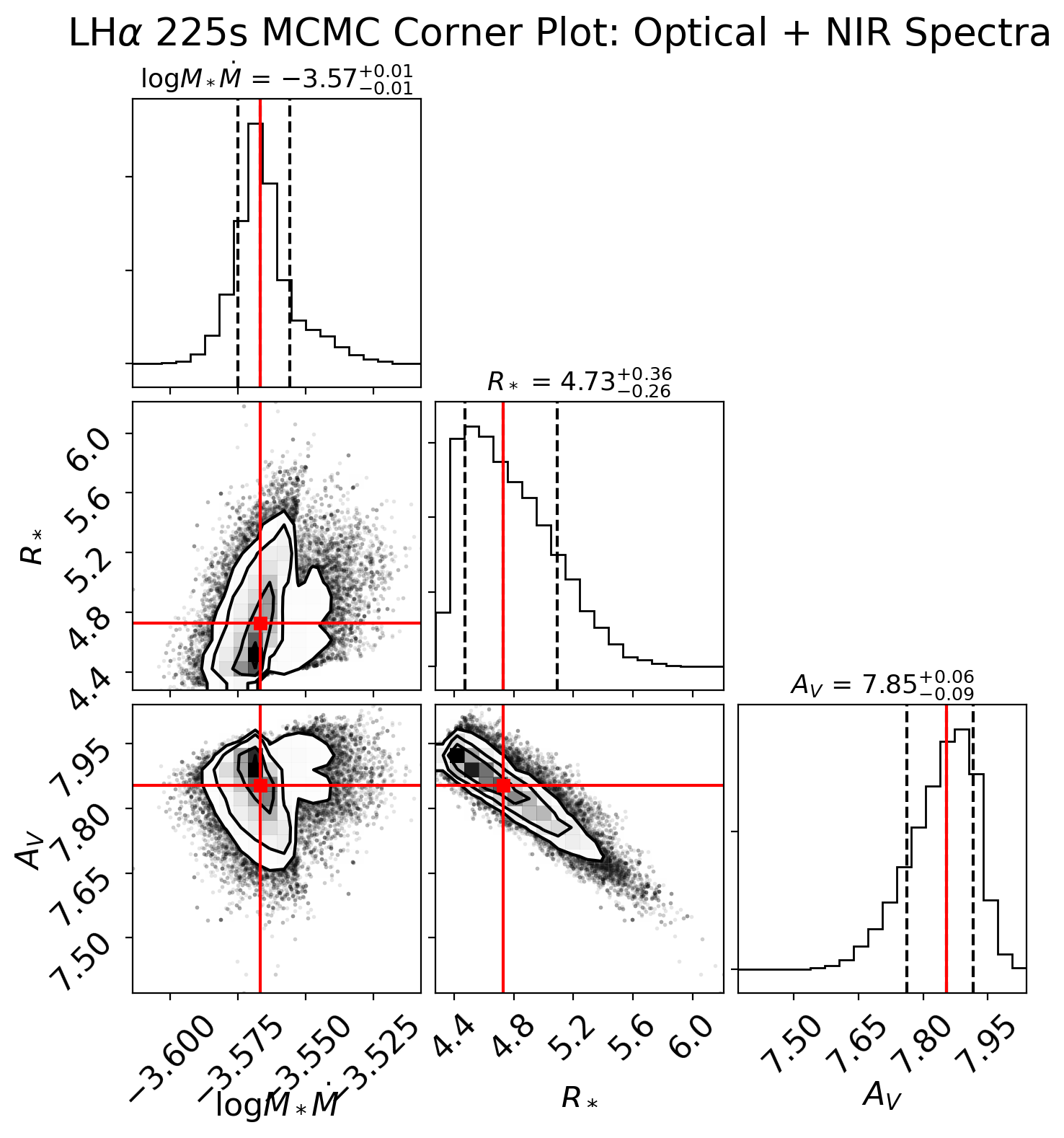}
\caption{
``Corner plot" output from the MC-MC fitting procedure applied to the spectral energy distribution of \lks\  
that is shown in Figure~\ref{fig:spec}.
The preferred parameter values are given above the boxes and
shown as red lines, with formal uncertainties indicated as vertical dashed lines.
The units in the panels are $M_\odot^2 ~yr^{-1}$, $R_\odot$, and magnitudes, along the diagonal.
}
\label{fig:mcmc}
\end{figure}

The SED that we model consists of the low resolution optical (Palomar/DBSP) and infrared (IRTF/SpeX) 
spectra, both of which are flux calibrated.  
Thus, they can be taken together as an accurate spectral energy distribution 
for the source.  As the observations were taken several months apart, 
the possibility of a flux offset between the optical and infrared SED needs to be considered.  
The date separation is only two periods (see Figure~\ref{fig:lczoomasas}), 
with just a 5\% offset in phase, and no obvious offset in flux;
thus we do not include any flux adjustment when fitting the SED.  We note that 
the observations correspond to a phase of the light curve 
near the minima of the $\sim$0.7 mag amplitude oscillations.

To sample the model parameter space and determine which values of the free parameters produce a model that adequately fits the data, we assume Gaussian likelihood and perform a Markov Chain Monte Carlo (MCMC) parameter estimation. We use MCMC to sample the non-analytic posterior probability distribution that is created once we develop the full accretion disk model, and have introduced physically reasonable priors for each of our given parameters. The Bayesian nature of MCMC also allows us to marginalize over all parameters except for a single one, and obtain credible intervals which constrain the value of a parameter given our model.

We use the {\it emcee} package to perform an affine-invariant MCMC routine to sample the parameter space. We explore both a uniform prior and the adoption of a Salpeter IMF for $M_*$ between $1 M_\odot$ and $15 M_\odot$, a uniform prior on $R_*$ between $0.3 R_\odot$ and $15 R_\odot$, a log-uniform prior on $\dot M$ between $10^{-7.5} M_\odot~yr^{-1}$ and  $10^{-3} M_\odot~yr^{-1}$, and a uniform prior on $A_V$ between 3.0 and 10.0. The log-uniform prior ensures that we sample accretion rates over the many orders of magnitude that are consistent with current FU Ori outburst theories.

We run the MCMC parameter explorations with 16 chains in our affine-invariant sampling, 4 for each free parameter. We run 6000 steps and take 50 percent of the steps as the burn-in period. Furthermore, to avoid autocorrelation in our MCMC techniques, we specify steps of $10^{-2}$ in the appropriate units for each parameter, in addition to our generous burn-in period. After we run our sampler, we test for convergence using the Gelman-Rubin statistic \citep{vats2018}, $\hat R$, and obtain the following values of $\hat R$ for the free parameters: $M_*: 1.191, R_*: 1.045, \dot M: 1.204, A_V: 1.036$. The higher values of the statistic for $M_*$ and $\dot M$ are to be expected, as these two parameters are strongly correlated in the corner plots, and do not converge to sharply peaked values individually. The $\hat R$ values for $R_*$ and $A_V$, however are below 1.1. If our MCMC iteration is to converge after infinite iterations, then $\hat R$ should converge to 1. We thus set a reasonable standard for $\hat R$, that it be in the range 1.001 to 1.5 \citep{vats2018}. Adopting a standard for convergence of $\hat R = 1.1$, then, allows us to say that the MCMC chains have converged adequately for $R_*$  and $A_V$.

The resulting values for the four free parameters and their uncertainties are: 
first for the flat IMF,
$M_* = 8.41^{+4.50}_{-4.21}~M_\odot$, 
$R_* = 4.75^{+0.36}_{-0.24}~R_\odot$, 
log $\dot M = -4.48^{+0.30}_{-0.19}$ dex $M_\odot~yr^{-1}$, 
and $A_V=7.87^{+0.06}_{-0.09}$ mag; 
and second for the Salpeter IMF,
$M_* = 2.75^{+4.85}_{-1.50}~M_\odot$, 
$R_* = 4.69^{+0.36}_{-0.25}~R_\odot$, 
log $\dot M = -4.00^{+0.34}_{-0.44}$ dex $M_\odot~yr^{-1}$, 
and $A_V=7.86^{+0.06}_{-0.09}$ mag. 
The accompanying corner plots for $M_*$ unfortunately resemble the assumed IMF,
rendering this parameter poorly constrained,
though it seems clear that $M_*$ must be at least several $M_\odot$ and possibly as high as $\sim 8 M_\odot$.
There is strong anti-correlation between $M_*$ and $\dot M$ as these two parameters trade off in shifting the SED peak. 
We further note that the likely more correct Salpeter IMF prior, produces a more poorly constrained $\dot M$ fit. 
Joint fitting, however, yields a much tighter constraint in the corner plots,
with log $(M_* \dot M) = -3.57^{+0.01}_{-0.01}$ dex $M^2_\odot~yr^{-1}$. 
This is consistent with the mean values quoted above for the individual parameters, 
even though their errors were much higher due to the degeneracy between them. 

$R_*$ and $A_V$ each have well-constrained fits in both the four-parameter and three-parameter fitting,
from which we conclude that their values are robust to assumptions about the IMF. 
These two parameters strongly control the overall shape of the SED. 
Although there is some anti-correlation between $R_*$ and $A_V$, because $A_V$ modifies the shape of the SED in a different way than $R_*$, their effects are decoupled by the MCMC parameter exploration, and their values are both meaningfully constrained as a result.
The model prediction of an extinction value $A_V= 7.9$ mag is consistent with 
our estimate based on the dereddening needed to match FGK spectral templates of
4-9 mag, and also the value of 7.2 mag found by \cite{m2019}.
The model prediction of a moderate mass, 
as well as a well-constrained radius $R= 4.7 R_\odot$
are consistent with previous indications that \lks\ is an intermediate-mass pre-main sequence star. 

In Figure~\ref{fig:mcmc} we show the results from the three-parameter ($M_* \dot M$ joint, $R_*$, and $A_V$) fitting.  
The luminosity in the accretion disk model is $L= 880 L_\odot$,
which compares well with the $L= 750 L_\odot$ estimated by \cite{m2019} from bright-state observations.
This luminosity is that required from accretion to match the optical/infrared spectral energy distribution,
and arises in the innermost $\sim 20~R_*$ of the disk;
it does not include energy re-radiated by the substantial amount of circumstellar dust in this source.
The maximum temperature in the accretion disk model is $T_{max} = 9400$ K.

\section{Discussion}

Our discussion is centered around three aspects of \lks:
the current bright-state quasi-periodicity, the presentation of 
a mixed-temperature absorption spectrum, and the overall interpretation
and context of this source in the panoply of young stellar object outbursts.

\subsection{Interpretation of the Quasi-Periodicity Near Light Curve Peak}

Periodicities in astrophysical sources usually find explanation in terms
of some form of rotational, orbital, or pulsational behavior.
In the case of \lks, there is no clear explanation.

The observed oscillatory behavior in \lks\ is non-sinusoidal but colorless. 
While the $g$ and $r$ magnitudes undergo $\sim 0.7$ mag periodic changes,
the $g-r$ color does not vary significantly.
We know that the oscillations are occurring now, and likely 
have been for the past $6+$ years of the current brightness maximum
of \lks\ -- given that \cite{m2019} also reported colorless variability, and
their photometry lines up fairly well with our phased light curve; see Figure~\ref{fig:phased}. 
However, there is no evidence that such oscillations were not 
occurring previously, in the data from earlier decades.  One line of evidence
against this is that, in contrast to the recent color-less oscillatory behavior, 
\cite{herbst1999} demonstrated a color-magnitude trend in the
brightness fluctuations of \lks, when it was around $15^m$, 
and color changes in $V-R$ of $\sim 3.5^m$ accompanying the brightness
drop of $\sim 4^m$. 

We can consider the observed large-amplitude periodicity of \lks\ relative to 
similar periodicities that have been detected in other young stellar objects.
These seem to fall into two categories\footnote{ 
Another set of accretion outburst sources with claimed low-amplitude 
periodicities are FU Ori, V2493 Cyg, and ASAS 13db.
In these objects, the evidence for truly periodic behavior -- as opposed to
mere ``time scales" for change -- seems somewhat weak.
In FU Ori, short-timescale (sub-day) variations were reported 
by \cite{kenyon2000} as non-periodic ``flickering" at the few percent level, 
and associated with the dynamical time at the inner edge of an accretion disk 
surrounding a $1~M_\odot, 4~R_\odot$ central source 
(in analogy to cataclysmic variable and x-ray binary systems).
\cite{errico2003}, \cite{herbig2003} and \cite{powell2012} all detected 
quasi-periodicities in line profiles of FU Ori, 
one at $\sim 13-15$ days associated with the wind, 
and one at $\sim 3.5$ days associated with the disk.
\cite{siwak2018} used high-precision photometry from $MOST$ to confirm  
photometric variations on approximately these same two time scales, finding
quasi-periodic signals at $\sim 10-11$ and $\sim 1.5-3$ days.
\cite{green2013} discussed a potential quasi-periodicity for 
V2493 Cyg (also known as PTF 10qpf and, in the pre-outburst stage, as HBC 722
or LkH$\alpha$ 188/G4), another FU Ori type object.  
Two timescales were also found for this object, 5.8 and 1.3 days. 
However, the phasing is not convincing, and
the findings have not been independently confirmed.
Finally, \cite{sa2017} argue for a $\sim 4$ day periodicity in the
outburst phases of ASAS 13db, an EX Lup type star, which they link to 
the rotation period of the star and attribute to the circulation of a new 
hot spot associated with the enhanced accretion event. Again, the phased 
light curves are not particularly compelling however.
}.
One features regular fluctuations of 0.5-2 mag
that are consistent over years to decades, and appear to indicate a cyclic
manner of accretion from a disk on to the central star. 
The other manifests only after a large-amplitude brightness increase, 
and appears as fluctuations superposed on top of a 
dominant, long-term accretion outburst.

V371 Ser \citep{hodapp2012,lee2020}, 
V2492 Cyg \citep{covey2011,hillenbrand2013},
V347 Aur \citep{dahm2020} 
all exhibit long timescale (months to $>1$ year) 
quasi-periodicities accompanied by color variations that have lasted for
perhaps decades.  The authors of the cited papers appeal
to some sort of cycling in the inner disk, perhaps driven by an unseen 
companion, as explanation for the observations.  In these sources, there is
a seemingly regular change between low-state and high-state accretion,
as evidenced by time series spectroscopy or color-magnitude behavior. 
These sources are distinct from EX Lup type objects, in that their brightenings 
are more regular, as opposed to more randomly timed, with light curve shapes
that are more rounded at their peaks, as opposed to the top-hat like
peaks that characterize e.g. EX Lup and V1647 Ori.

More similar to the case of \lks\ than the above regularly varying
accretors, are sources like L1634 IRS7 and ESO $H\alpha$-99. These 
have both shown brightness quasi-periodicity following a 
clearly detected, long-timescale, large-amplitude outburst.  
\cite{hodapp2015} demonstrated L1634 IRS7 to have a remarkable $\sim$2 mag 
amplitude at $K_s$-band, with a 37 day periodic signal, that is
superposed on a slow rise over about 2 decades.  
\cite{hodapp2019} classified ESO $H\alpha$-99 as an EX Lup type star, 
but one that exhibits $\approx$0.5 mag amplitude variations 
on timescales of about a month, with a semi-periodic morphology.
Again the behavior is superposed on a brightness plateau that followed
a 4.4 mag brightening over about a year.  
The ESO $H\alpha$-99 brightness oscillations are interpreted
in terms of a rotating structure within the circumstellar disk,
an explanation we also explore for the situation of \lks.

Like L1634 IRS7 and ESO $H\alpha$-99,
\lks\ has a quasi-periodic light curve morphology during its light curve 
plateau, and the amplitudes are also comparable 
at about a factor of two fluctuations.  However, \lks\ has experienced a 
much larger-amplitude overall brightening ($>$7 mag), 
on a significantly longer time scale ($\sim$15 years).
If orbital in origin, 
the 43 day period in \lks\ would correspond to changes in brightness 
occurring on a size scale of $\sim 0.49 \times (M/8.4~M_\odot)$ AU.
This is still inside the outbursting disk according to our disk model (\S7),
in a region where the temperature of the disk is $\approx$1750 K.

The temperature range is interesting, because it corresponds to 
the approximate temperature where the TiO and VO molecules that we observe
start to exist ($1500 \gtrapprox T \lessapprox 4000$ K). It is also consistent with
the dust destruction temperature ($T \gtrapprox 1200-1800$ K), and thus could 
coincide with a region where the hot disk is cool enough to also have dust.
The periodicity, if driven by orbital phenomena, may be related
to turbulence or to azimuthally asymmetric phenomena occurring at this 
special location.
However, the periodic variability we observe is in the optical
($g$ and $r$ bands). Here, the disk is expected to be much hotter
than the relevant $\approx$1750 K. Specifically, the photons dominating the flux
at these wavelengths are coming from $\approx$7000-7400 K regions 
in our disk model.  It is unclear what in this temperature range
could be varying on the observed 43 day time scale.

In the simulations of \cite{zhu2009}, a gravitational instability somewhat
further out in the disk causes matter to rush inward, triggering 
a magneto-rotational instability, which in turn triggers a thermal instability
in the innermost disk. The temperature at which the magneto-rotational instability
begins to take effect is $\sim$1200 K, somewhat lower than the $\approx$1750 K
estimate above. Lower temperatures are possible for outburst accretion rates that are
lower than the fiducial model of $10^{-4} M_\odot~yr^{-1}$. 



An alternate to being a special place in the disk, as argued above, 
is that the periodic variability near maximum brightness for \lks, 
is related to an orbiting companion, that affects the disk brightness 
quasi-periodically, on its orbital timescale.  As above, 
given our estimate of $M_*$ and the inferred period, the purported companion 
would be located at 0.49-0.62 AU, depending on the mass ratio. 
The current bright state yields no evidence in photometry or
radial velocities for a binary companion. However,
if the period is associated with an unseen companion located within the
disk instability region, that companion might be acting as 
a gatekeeper for accretion from the outer disk, causing the
quasi-periodic modulation of the accretion by a factor of two (0.7 mag).
It is unclear whether the profile of the light curve (Figure~\ref{fig:phased}) 
is consistent with such a scenario, however.

Another possibility is some sort of slow pulsational behavior. 
One scenario is pulsational ringing following a planet engulfment.  
In order to produce a period of 43 days, however, 
the star would have to have swelled to 230 $R_\odot$, 
which is $\sim$1.5 dex larger than the size inferred from our SED modelling,
and therefore seems unlikely.
Another pulsational scenario is that of acoustic mode trapping in the
disk-to-star boundary layer and the inner disk, 
studied by \cite{belyaev2012}.   Both the amplitudes and periods 
are much smaller (Coleman \& Ravikov, in preparation) than in our case, however.

\subsection{Interpretation of The Mixed Absorption Spectrum}

Summarizing the inferences that can be made from the object's spectrum,
we have found evidence for a mid-A spectral type from the high-resolution
optical spectrum, an FGK spectral type from the low resolution
optical spectrum, and also absorption from much cooler H$_2$O molecules 
in the infrared spectrum, as might characterize an M spectral type.
At a minimum, we can conclude that the spectral type is mixed, or composite. 
 We further
offer the hypothesis that what we may be seeing in this source is a disk
spectrum, but with a warm wind that gives rise to the hotter spectral type
inferred from the optical metal lines in the high dispersion data.

We developed the disk hypothesis in more detail in \S7\ by fitting the optical to near-infrared SED.
The stellar mass derived from the disk modelling is
several $M_\odot$ and possibly as high as $\sim 8 M_\odot$, which 
would correspond to a spectral type later than B2 on the main sequence, 
and temperature $<20,500$ K.   
The maximum temperature in the innermost disk annulus is
$9400$ K in the model, which would correspond to a spectral type of A1.  
Rather than seeing the hottest parts of the disk directly, the
spectral type we estimate from the optical high dispersion spectrum seems
dominated by a component that is a bit later, $\sim$A3-A6. 

The seemingly disparate pieces of information above are consistent with one 
another in the scenario where an embedded young star has undergone
an outburst, and is now presenting a disk-dominated spectrum.
In addition to the evidence for a mixed or composite absorption spectrum,
we have reported that 
the absorption lines in \lks\ have the narrow width of V1515 Cyg
(supporting a low inclination), but the relatively shallow depth of V1057 Cyg
(supporting significant disk broadening). 
A comparison to FU Ori itself also shows good agreement, with the exception of
the lower excitation lines which are in absorption in FU Ori, 
but in emission in \lks.

\subsection{ What is this source?}

\lks\ has brightened by $>$7 mag over the past two decades,
seemingly without much change in its optical color, but a clear blue-ing
of the near-infrared spectral energy distribution.
The color changes imply that extinction clearing is not a dominant factor
in the brightening.  Instead, the behavior is more consistently explained
by the effects of enhanced accretion, or a combination of enhanced
accretion and extinction clearing.

\cite{connelley2018} state, based on the archetypal examples of FU Ori,
V1057 Cyg, and V1515 Cyg, that FU Ori stars are spectroscopically defined as
having (in infrared spectra):
``strong CO absorption, weak metal absorption, strong water bands, low gravity, strong blueshifted He I absorption, and few (if any) emission lines".
By this definition, \lks\ does not qualify as an FU Ori star.
Yet it does have FU Ori-like absorption in the blue optical 
(Figure~\ref{fig:blueandtio},~\ref{fig:hires}), though
without much indication of a systematic spectral type
dependence on wavelength (Figure~\ref{fig:templates}).
Instead, the dominant spectral class is fairly consistent across wavelengths
at $\sim$ mid-A, though there is also reasonable correlation in some parts
of the spectrum with cooler FG and early K temperatures, and in the infrared,
with M-type temperatures.  The source also has low gravity signatures 
in the form of \ion{Sr}{2} and \ion{Ba}{2} lines.


The weak emission line spectrum of \lks\ is similar to, but more muted than
that of other outbursting young stars such as Gaia 19ajj, V2492 Cyg (PTF 10nvg),
and V1647 Ori.  Gaia 19ajj has only atomic emission,
which is slightly stronger than what is seen in \lks.
V2492 Cyg has similar atomic but also the molecular emission in TiO/VO and CO
that is seen in \lks; it differs from \lks\ in having H$_2$O emission, 
which is in absorption in \lks\ (as well as in Gaia 19ajj).
Both V2492 Cyg (PTF 10nvg) and V1647 Ori have stronger
emission lines and much less of an absorption signature than seen in
the mixed-temperature spectrum of \lks.

We support the conclusion of \cite{m2019} that there is a strong wind 
in the outburst state of \lks\ (Figure~\ref{fig:wind},~\ref{fig:nirspec}).
The relatively low wind velocity, only about $-150$ \kms, suggests a low to
modest inclination for the source.
A low inclination is also supported by the narrow absorption line widths
in the optical.  Further, the shape of the CO bandhead lines lacks the morphology
expected \citep{Martin1997} from a highly inclined source.

Overall, \lks\ in outburst seems a lot like Gaia 19ajj, sharing many of
the properties described above.  
Both sources have evidence for large-amplitude optical outbursts, taking
years to more than a decade to develop, and previous bright photometric states,
with potential repeat timescales of one to a few decades.
\lks\ has a stronger absorption 
spectrum that is more similar to that in the FU Ori stars at the bluest wavelengths. 
It also has TiO emission, which was not seen in Gaia 19ajj,
as well as stronger CO emission.  \lks\ has a weaker emission spectrum
than Gaia 19ajj, that is present only at red optical and near-infrared wavelengths.  


A further interesting aspect of \lks\ is its mass, which is likely
in the range of several to $\sim 8~M_\odot$.  As such, while not massive,
it certainly qualifies as an intermediate-mass young star undergoing accretion-driven outburst behavior.

\section{Conclusions}

We have reported on a detailed study of the spectroscopic and
photometric presentation of \lks\ during its current bright state.
Our study
adds to the work of \cite{m2019}, who first reported the dramatic 
brightening of this enigmatic young stellar object.
\lks\ was visible in the 1950's (POSS-I plates), but not in the
early 1980s (Quick-V plates). It was recorded as moderately bright in the 
late 1980s \citep{shevchenko1993}, then faded by the 1990's (POSS-II plates,
\citet{hillenbrand1995,herbst1999}). It appears to have started brightening
again in the late 2000s, reaching something of a plateau in $\sim$ 2015.

Our main new findings regarding \lks\ are:
\begin{itemize}
\item
A non-sinusoidal, colorless, quasi-periodic light curve in the bright state. 
While the phased light curve has significant dispersion,
the oscillations have persisted for over 3 years.  
The derived period is $\sim 43$ days and the amplitude $\sim 0.7$ mag.
\item
A currently decreasing emission spectrum 
and increasing absorption spectrum, based on limited spectral time series data
as the outburst continues to develop in its long plateau phase.
\item
Absorption features at blue optical wavelengths and in the near-infrared
which indicate a mixed-temperature photosphere having mid-A to M-type
absorption components present. The spectrum can not be characterized 
with a single spectral type in the traditional sense.
Low-gravity is evidenced by the strength of \ion{Sr}{2} and \ion{Ba}{2} lines
\item
Strong wind signatures in $H\alpha$, \ion{He}{1} 10830 \AA,
and a variety of metal lines including \ion{Ca}{2} doublet and triplet, 
\ion{Fe}{2}, \ion{Na}{1}D, \ion{K}{1}, \ion{Mg}{1}b, \ion{O}{1} triplet, 
and \ion{Li}{1}.
\item
Atomic emission lines that are strongest in the red optical and
short-wavelength infrared, dominating the spectrum at these wavelengths,
but essentially absent at blue optical wavelengths.
\item 
Molecular emission from TiO and CO both before and after the brightening,
with no evidence for line-to-continuum change.
Molecular emission from H$_2$ that, relative to the continuum, weakened
after the brightening.
\item 
Molecular absorption by H$_2$O that was enhanced after the brightening.
\item
As the star brightened,
a decrease in the steep spectral slope, or blue-ing behavior corresponding to 
a reduction in extinction by $A_V \approx 3$ mag.  We find a current 
extinction value of $A_V \approx 8$ mag.
\item
Stellar parameters of $M_* =$ several to 8 $M_\odot$ and $R_* = 4.7~R_\odot$, 
and an accretion rate of log $\dot M = -4.5$ to $-4$ dex $M_\odot~yr^{-1}$ 
result from an accretion disk model fit to flux-calibrated spectrophotometry.  
The corresponding luminosity in the accretion disk model is $L = 880~L_\odot$.
\item
Evidence for a relatively low source inclination, based on (disk-broadened) absorption line widths, 
CO bandhead emission profiles, and P-Cygni profile terminal velocities, 
consistent with inferences in previous literature.
\end{itemize}

\lks\ deserves further study, but the current best interpretation is that
the source has undergone a long-timescale accretion outburst, perhaps
related to an orbiting companion at $\approx 0.49$ AU.
Phenomenologically, the source bears some resemblance to FU Ori stars
in the blue wavelengths, with a mixed-temperature pure-absorption
absorption spectrum.  
\lks\ is not entirely FU Ori-like though, considering the weak low-excitation 
metal emission and the lack of strong molecular absorption in the near-infrared.
It is like PTF 14jg in this regard, though not as hot. 

The large-amplitude, very long-timescale photometric variations of \lks\ 
are like those of sources such as PV Cep, V2492 Cyg (PTF 10nvg), and Gaia 19ajj.
However, while these sources share aspects of their emission-line spectra 
at red optical and near-infrared wavelengths (which are even stronger than in \lks), 
they do not have prominent absorption spectra like \lks\ does.
The quasi-periodicity of \lks\ near its current light curve peak 
is reminiscent of the infrared sources L1634 IRS7 and ESO $H\alpha$-99, 
which are also similar in exhibiting near-infrared CO emission.

\facility{IPHaS, PanSTARRS, AAVSO, Gaia, PO:1.2m(PTF), PO:1.2m(ZTF), IRSA, ASAS, PO:Hale(DBSP), Keck:I(HIRES), Keck:II(NIRSPEC), IRTF(SpeX), Gemini(NIRI)}

\vskip0.25truein
\begin{acknowledgments}
We acknowledge the work of former Caltech undergraduate Sirin Caliskan, who
produced the photometry and astrometry for the Keck/LRIS images. 
We acknowledge with thanks the AAVSO International Database which allowed us
to connect with co-author DRP.
We thank Roc Cutri for consultation regarding whether the saturated
photometry of \lks\ could be recovered in NEOWISE data.
We thank Kishaley De for checking on the source in Gattini data.
Andrew Howard facilitated the first Keck/HIRES spectrum.
Erik Petigura and Trevor David kindly obtained the first Keck/NIRSPEC data set;
Jessica Spake kindly allowed for the second.
We thank Richard Larson for his interest in this system,
and for his suggestions regarding the 43 day quasi-periodicity.
We also thank the referee for a careful and insightful review of our work.
\end{acknowledgments}


\vskip0.25truein


\begin{thebibliography}


\bibitem[Andrillat \& Swings(1976)]{andrillat1976} Andrillat, Y., \& Swings, J.~P.\ 1976, \apjl, 204, L123

\bibitem[Aspin et al.(1994)]{aspin1994} Aspin, C., Sandell, G., \& Weintraub, D.~A.\ 1994, \aap, 282, L25 

\bibitem[Bae et al.(2011)]{bae2011} Bae, J.-H., Kim, K.-T., Youn, S.-Y., et al.\ 2011, \apjs, 196, 21. 

\bibitem[Barentsen et al.(2013)]{barentsen2013} Barentsen, G., Vink, J.~S., Drew, J.~E., \& Sale, S.~E.\ 2013, \mnras, 429, 1981 

\bibitem[Bautista et al.(2015)]{Bautista2015} Bautista, M.~A., Fivet, V., Ballance, C., et al.\ 2015, \apj, 808, 174. 
 

\bibitem[Bellm et al.(2019)]{bellm2019} Bellm, E.~C., Kulkarni, S.~R., Graham, M.~J., et al.\ 2019, \pasp, 131, 018002 

\bibitem[Belyaev et al.(2012)]{belyaev2012} Belyaev, M.~A., Rafikov, R.~R., \& Stone, J.~M.\ 2012, \apj, 760, 22. 

\bibitem[Carvalho \& Hillenbrand(2021)]{Carvalho2021} Carvalho, A. \& Hillenbrand, L.A., 2021, AAS Journals, submitted

\bibitem[Chubak et al.(2012)]{chubak2012} Chubak, C., Marcy, G., Fischer, D.~A., et al.\ 2012, arXiv e-prints, arXiv:1207.6212


\bibitem[Connelley \& Reipurth(2018)]{connelley2018} Connelley, M.~S., \& Reipurth, B.\ 2018, \apj, 861, A145 

\bibitem[Covey et al.(2011)]{covey2011} Covey, K.~R., Hillenbrand, L.~A., Miller, A.~A., et al.\ 2011, \aj, 141, 40

\bibitem[Cutri et al.(2003)]{cutri2003} Cutri, R.~M., Skrutskie, M.~F., van Dyk, S., et al.\ 2003, ``The IRSA 2MASS All-Sky Point Source Catalog, NASA/IPAC Infrared Science Archive. 
{\url{http://irsa.ipac.caltech.edu/applications/Gator/}}


\bibitem[Cutri et al.(2012)]{cutri2012viz} Cutri, R.~M., et al.\ 2012, VizieR Online Data Catalog, 2311 (WISE)
\bibitem[Cutri et al.(2012)]{cutri2012} Cutri, R.~M., Wright, E.~L., Conrow, T., et al.\ 2012, Explanatory Supplement to the WISE All-Sky Data Release Products


\bibitem[Cutri \& et al.(2014)]{cutri2014viz} Cutri, R.~M., \& et al.\ 2014, VizieR Online Data Catalog, II/328 (AllWISE)
\bibitem[Cutri et al.(2013)]{cutri2013} Cutri, R.~M., Wright, E.~L., Conrow, T., et al.\ 2013, Explanatory Supplement to the AllWISE Data Release Products, by R. M. Cutri et al.


\bibitem[Cutri et al.(2015)]{cutri2015} Cutri, R.~M., Mainzer, A., Conrow, T., et al.\ 2015, Explanatory Supplement to the NEOWISE Data Release Products, 
{\url{http://wise2.ipac.caltech.edu/docs/release/neowise/expsup}}

\bibitem[Dahm \& Hillenbrand (2020)]{dahm2020} Dahm, S.E. \& Hillenbrand, L.A.\ 2020, AJ, 160, 278

\bibitem[De et al.(2020)]{de2020} De, K., Hankins, M. J., Kasliwal, M. M., et al. 2020, PASP, 132, 025001

\bibitem[Errico et al.(2003)]{errico2003} Errico, L., Vittone, A., \& Lamzin, S.~A.\ 2003, Astronomy Letters, 29, 105

\bibitem[Fitzpatrick(1999)]{Fitzpatrick1999} Fitzpatrick, E.~L.\ 1999, \pasp, 111, 63. 
 

\bibitem[Flewelling et al.(2020)]{flewelling2020} Flewelling, H.~A., Magnier, E.~A., Chambers, K.~C., et al.\ 2020, \apjs, 251, 7. 

\bibitem[Gaia Collaboration et al.(2018)]{gaiadr2} Gaia Collaboration, Brown, A.~G.~A., Vallenari, A., et al.\ 2018, \aap, 616, A1

\bibitem[Graham et al.(2019)]{graham2019} Graham, M.~J., Kulkarni, S.~R., Bellm, E.~C., et al.\ 2019, \pasp, 131, 078001 

\bibitem[Green et al.(2013)]{green2013} Green, J.~D., Robertson, P., Baek, G., et al.\ 2013, \apj, 764, 22

\bibitem[Gutermuth et al.(2009)]{gutermuth2009} Gutermuth, R.~A., Megeath, S.~T., Myers, P.~C., et al.\ 2009, \apjs, 184, 18

\bibitem[Hankins et al.(2020)]{hankins2020} Hankins, M., Hillenbrand, L.~A., De, K., et al.\ 2020, The Astronomer's Telegram, 13902

\bibitem[Hauschildt et al.(1999)]{Hauschildt1999} Hauschildt, P.~H., Allard, F., Ferguson, J., et al.\ 1999, \apj, 525, 871. 

\bibitem[Heinze et al.(2018)]{heinze2018} Heinze, A.~N., Tonry, J.~L., Denneau, L., et al.\ 2018, \aj, 156, 241. 


\bibitem[Herbig(1960)]{herbig1960} Herbig, G.~H.\ 1960, \apjs, 4, 337

\bibitem[Herbig et al.(2003)]{herbig2003} Herbig, G.~H., Petrov, P.~P., \& Duemmler, R.\ 2003, \apj, 595, 384

\bibitem[Herbst \& Shevchenko(1999)]{herbst1999} Herbst, W., \& Shevchenko, V.~S.\ 1999, \aj, 118, 1043

\bibitem[Herczeg \& Hillenbrand(2014)]{herczeg2014} Herczeg, G.~J., \& Hillenbrand, L.~A.\ 2014, \apj, 786, 97

\bibitem[Hillenbrand et al.(2019c)]{hillenbrand2019c} Hillenbrand, L.~A., The Astronomer's Telegram, 13321, 1

\bibitem[Hillenbrand et al.(1995)]{hillenbrand1995} Hillenbrand, L.~A., Meyer, M.~R., Strom, S.~E., et al.\ 1995, \aj, 109, 280

\bibitem[Hillenbrand et al.(2012)]{hillenbrand2012} Hillenbrand, L.~A., Knapp, G.~R., Padgett, D.~L., et al.\ 2012, \aj, 143, 37


\bibitem[Hillenbrand et~al.(2013)]{hillenbrand2013} Hillenbrand, L.~A., Miller, A.~A., Covey, K.~R., et~al.\ 2013, \aj, 145, 59

\bibitem[Hillenbrand et al.(2018)]{hillenbrand2018} Hillenbrand, L.~A., Contreras Pe{\~n}a, C., Morrell, S., et al.\ 2018, \apj, 869, 146 

\bibitem[Hillenbrand et al.(2019b)]{hillenbrand2019b} Hillenbrand, L.~A., Reipurth, B., Connelley, M., et al.\ 2019, \aj, 158, 240

\bibitem[Hillenbrand et al.(2019a)]{hillenbrand2019a} Hillenbrand, L.~A., Miller, A.~A., Carpenter, J.~M., et al.\ 2019, \apj, 874, 82

\bibitem[Hodapp et al.(1996)]{hodapp1996} Hodapp, K.W., ApJ, 468, 861. 

\bibitem[Hodapp et al.(2012)]{hodapp2012} Hodapp, K. W., Chini, R., Watermann, R., \& Lemke, R. 2012, ApJ, 744, 56

\bibitem[Hodapp, \& Chini(2015)]{hodapp2015} Hodapp, K.~W., \& Chini, R.\ 2015, \apj, 813, 107

\bibitem[Hodapp et al.(2019)]{hodapp2019} Hodapp, K.~W., Reipurth, B., Pettersson, B., et al.\ 2019, \aj, 158, 241

\bibitem[{{Howard} {et~al.}(2010){Howard}, {Johnson}, {Marcy}, {Fischer},
  {Wright}, {Bernat}, {Henry}, {Peek}, {Isaacson}, {Apps}, {Endl}, {Cochran},
  {Valenti}, {Anderson}, \& {Piskunov}}]{howard2010}
{Howard}, A.~W., {Johnson}, J.~A., {Marcy}, G.~W., {et~al.} 2010, \apj, 721,
  1467

\bibitem[Ibragimov et al.(1988)]{ibragimov1988} Ibragimov, M.~A., Mel'Nikov, S.~Y., Chernyshov, A.~V., et al.\ 1988, Astrofizika, 29, 633

\bibitem[Jordi et al.(2006)]{jordi2006} Jordi, K., Grebel, E.~K., \& Ammon, K.\ 2006 \aap, 460, 339 

\bibitem[{{Kenyon} {et~al.}(1988){Kenyon}, {Hartmann}, \&
  {Hewett}}]{kenyon1988}
{Kenyon}, S.~J., {Hartmann}, L., \& {Hewett}, R. 1988, \apj, 325, 231

\bibitem[Kenyon et al.(2000)]{kenyon2000} Kenyon, S.~J., Kolotilov, E.~A., Ibragimov, M.~A., et al.\ 2000, \apj, 531, 1028

\bibitem[{{Lee} {et~al.}(2020){Lee}, {Lee}, {Aikawa}, {Herczeg}, \& {Johnstone}}]{lee2020} {Lee}, S., {Lee}, J.-E., {Aikawa}, Y., {Herczeg}, G., \& {Johnstone}, D. 2020, \apj, 889, 20


\bibitem[Looney et al.(2006)]{looney2006} Looney, L.~W., Wang, S., Hamidouche, M., et al.\ 2006, \apj, 642, 330

\bibitem[Magakian \& Movsessian(1997)]{m1997} Magakian, T.~Y. \& Movsessian, T.~A.\ 1997, Astronomy Letters, 23, 666

\bibitem[Magakian et al.(2019)]{m2019} Magakian, T.~Y., Movsessian, T.~A., Andreasyan, H.~R., et al.\ 2019, \aap, 625, A13

\bibitem[Maksimova et al.(2020)]{maksimova2020A} Maksimova, L.~A., Pavlyuchenkov, Y.~N., \& Tutukov, A.~V.\ 2020, Astronomy Reports, 64, 815. 

\bibitem[Martin(1997)]{Martin1997} Martin, S.~C.\ 1997, \apjl, 478, L33. 

\bibitem[Martin et al.(2018)]{martin2018} Martin, E.~C., Fitzgerald, M.~P., McLean, I.~S., et al.\ 2018, Ground-based and Airborne Instrumentation for Astronomy VII, \procspie, 10702, 107020A 

\bibitem[Marvel(2005)]{marvel2005} Marvel, K.~B.\ 2005, \aj, 130, 2732

\bibitem[Masci et al.(2019)]{masci2019} Masci, F.~J., Laher, R.~R., Rusholme, B., et al.\ 2019, \pasp, 131, 018003 

\bibitem[Matthews et al.(2007)]{matthews2007} Matthews, B.~C., Graham, J.~R., Perrin, M.~D., et al.\ 2007, \apj, 671, 483

\bibitem[McLean et al.(1998)]{mclean1998} McLean, I.S., Becklin, E.E., Bendiksen, O., et al.\ 1998, \procspie, 3354, 566

\bibitem[Miller et al.(2015)]{miller2015} Miller, A.~A., Hillenbrand, L.~A., Bilgi, P., et al.\ 2015, The Astronomer's Telegram, 7428

\bibitem[Mora et al.(2001)]{mora2001} Mora, A., Mer{\'\i}n, B., Solano, E., et al.\ 2001, \aap, 378, 116

\bibitem[Navarete et al.(2015)]{navarete2015} Navarete, F., Damineli, A., Barbosa, C.~L., et al.\ 2015, \mnras, 450, 4364. 

\bibitem[Oke et al.(1995)]{oke1995} Oke, J.~B., Cohen, J.~G., Carr, M., et al.\ 1995, \pasp, 107, 375

\bibitem[Oke \& Gunn(1982)]{og1982} Oke, J. B., \& Gunn, J. E. 1982, PASP, 94, 586

\bibitem[Palla et al.(1995)]{palla1995} Palla, F., Testi, L., Hunter, T.~R., et al.\ 1995, \aap, 293, 521

\bibitem[Pecchioli et al.(2016)]{Pecchioli2016} Pecchioli, T., Sanna, N., Massi, F., et al.\ 2016, \pasp, 128, 073001. 


\bibitem[Petrov \& Herbig(2008)]{petrov2008} Petrov, P.~P., \& Herbig, G.~H.\ 2008, \aj, 136, 676

\bibitem[Powell et al.(2012)]{powell2012} Powell, S.~L., Irwin, M., Bouvier, J., et al.\ 2012, \mnras, 426, 3315

\bibitem[Prugniel et al.(2007)]{prugniel2007} Prugniel, P., Soubiran, C., Koleva, M., et al.\ 2007, arXiv e-prints, astro-ph/0703658

\bibitem[Purser et al.(2021)]{purser2021} Purser, S.~J.~D., Lumsden, S.~L., Hoare, M.~G., et al.\ 2021, \mnras, 504, 338. 


\bibitem[Rayner et al.(2003)]{rayner2003} Rayner, J.~T., Toomey, D.~W., Onaka, P.~M., et al.\ 2003, \pasp, 115, 362 

\bibitem[Rodriguez \& Hillenbrand(2022)]{rodriguez2022} Rodriguez, A.~C. \& Hillenbrand, L.~A.\ 2022, arXiv:2112.01549

\bibitem[Sandell et al.(2012)]{sandell2012} Sandell, G., Wiesemeyer, H., Requena-Torres, M.~A., et al.\ 2012, \aap, 542, L14. 

\bibitem[{{Shappee} {et~al.}(2014){Shappee}, {Prieto}, {Stanek}, {Kochanek},
  {Holoien}, {Jencson}, {Basu}, {Beacom}, {Szczygiel}, {Pojmanski},
  {Brimacombe}, {Dubberley}, {Elphick}, {Foale}, {Hawkins}, {Mullins},
  {Rosing}, {Ross}, \& {Walker}}]{shappee2014}
{Shappee}, B., {Prieto}, J., {Stanek}, K.~Z., {et~al.} 2014, in American
  Astronomical Society Meeting Abstracts, Vol. 223, American Astronomical
  Society Meeting Abstracts \#223, 236.03

\bibitem[{{Sharon} {et~al.}(2010){Sharon}, {Hillenbrand}, {Fischer}, \&
  {Edwards}}]{sharon2010}
{Sharon}, C., {Hillenbrand}, L., {Fischer}, W., \& {Edwards}, S. 2010, \aj, 139, 646

\bibitem[Shevchenko et al.(1993)]{shevchenko1993} Shevchenko, V.~S., Grankin, K.~N., Ibragimov, M.~A., et al.\ 1993, \apss, 202, 121

\bibitem[Sicilia-Aguilar et al.(2017)]{sa2017} Sicilia-Aguilar, A., Oprandi, A., Froebrich, D., et al.\ 2017, \aap, 607, A127

\bibitem[Siwak et al.(2018)]{siwak2018} Siwak, M., Winiarski, M., Og{\l}oza, W., et al.\ 2018, \aap, 618, A79

\bibitem[Strom et al.(1972)]{strom1972} Strom, K.~M., Strom, S.~E., Breger, M., et al.\ 1972, \apjl, 173, L65

\bibitem[Terranegra et al.(1994)]{terranegra1994} Terranegra, L., Chavarria-K., C., Diaz, S., et al.\ 1994, \aaps, 104, 557

\bibitem[van den Ancker et al.(2000)]{vandenancker2000} van den Ancker, M.~E., Wesselius, P.~R., \& Tielens, A.~G.~G.~M.\ 2000, \aap, 355, 194

\bibitem[VanderPlas(2018)]{vdp2018} VanderPlas, J.~T.\ 2018, \apjs, 236, 16

\bibitem[Vats \& Knudson(2018)]{vats2018} Vats, D., \& Knudson, C.\ 2018, arXiv e-prints, arXiv:1812.09384

\bibitem[Vogt et al.(1994)]{vogt1994} Vogt, S.~S., Allen, S.~L., Bigelow, B.~C., et al.\ 1994, \procspie, 2198, 362 


\bibitem[Wenzel(1972)]{wenzel1972} Wenzel, W.\ 1972, Information Bulletin on Variable Stars, 713, 1

\bibitem[{{Zhu} {et~al.}(2007){Zhu}, {Hartmann}, {Calvet}, {Hernand ez},
  {Muzerolle}, \& {Tannirkulam}}]{zhu2007}
{Zhu}, Z., {Hartmann}, L., {Calvet}, N., {et~al.} 2007, \apj, 669, 483

\bibitem[{{Zhu} {et~al.}(2009){Zhu}, {Hartmann}, {Gammie}, \&
  {McKinney}}]{zhu2009}
{Zhu}, Z., {Hartmann}, L., {Gammie}, C., \& {McKinney}, J.~C. 2009, \apj, 701,
  620

\bibitem[{{Zhu} {et~al.}(2020){Zhu}, {Jiang}, \& {Stone}}]{zhu2020}
{Zhu}, Z., {Jiang}, Y.-F., \& {Stone}, J.~M. 2020, \mnras, 495, 3494

\end{thebibliography}

\end{document}